\documentclass[%
 reprint,
superscriptaddress,
nofootinbib,
 amsmath,amssymb,
 aps,
prb,
]{revtex4-2}

\usepackage{graphicx}
\usepackage{dcolumn}
\usepackage{bm}
\usepackage{bbold}
\usepackage{soul}

\usepackage{tikz}

\def\dimer{\tikz[baseline=-0.5ex]{
			\fill (0,0) circle (1.5pt) coordinate (A);
			\fill (3.0ex, 0) circle (1.5pt) coordinate (B);
			\draw[black, thick] (A)--(B);}
	}

\def\entangleddota{\tikz[baseline=-0.5ex]{
        \fill (0,0) circle (1.5pt) coordinate (A);
        \fill (3.0ex, 0) circle (1.5pt) coordinate (B);
        \draw[black, thick, dotted] (A)--(B);}
}

\def\entangleddimerdot{\tikz[baseline=-0.5ex]{
		\fill (0,0) circle (1.5pt) coordinate (A);
		\fill (3.0ex, 0) circle (1.5pt) coordinate (B);
		\fill (6.0ex,0) circle (1.5pt) coordinate (C);
		\fill (9.0ex, 0) circle (1.5pt) coordinate (D);
		\draw[black, thick] (B)--(C);
		\draw[black, thick, dotted] (A) .. controls (3.0ex, 1.5ex) and (6.0ex, 1.5ex) .. (D);
	}
}

 \def\onedimera{\tikz[baseline=-0.5ex]{
			\fill (0,0) circle (1.5pt) coordinate (A);
			\fill (3.0ex, 0) circle (1.5pt) coordinate (B);
			\draw[black, thick] (A)--(B);
			\fill (6.0ex,0) circle (1.5pt) coordinate (C);
			\fill (9.0ex, 0) circle (1.5pt) coordinate (D);
			\node at (2.8ex, -1.6ex) {\scriptsize $j$};
			\node at (6.3ex, -1.6ex) {\scriptsize $j+1$};
            \draw[black, thick, dotted] (C)--(D);
		}
	}

\def\redbluetwodimer{\tikz[baseline=-0.5ex]{
		\fill[red] (0,0) circle (1.5pt) coordinate (A);
		\fill[blue] (3.0ex, 0) circle (1.5pt) coordinate (B);
		\fill[blue] (6.0ex,0) circle (1.5pt) coordinate (C);
		\fill[red] (9.0ex, 0) circle (1.5pt) coordinate (D);
		\draw[blue, thick] (B)--(C);
		\draw[red, thick] (A) .. controls (3.0ex, 1.5ex) and (6.0ex, 1.5ex) .. (D);
	}
}

\def\onedot{\tikz[baseline=-0.5ex]{
			\fill (0,0) circle (1.5pt) coordinate (A);
		}
	}

\def\twodota{\tikz[baseline=-0.5ex]{
			\fill[blue] (0,0) circle (1.5pt) coordinate (A);
			\fill[red] (3.0ex, 0) circle (1.5pt) coordinate (B);
		}
	}
	
\def\twodotb{\tikz[baseline=-0.5ex]{
			\fill[red] (0,0) circle (1.5pt) coordinate (A);
			\fill[blue] (3.0ex, 0) circle (1.5pt) coordinate (B);
		}
	}

 \def\reddimer{\tikz[baseline=-0.5ex]{
			\fill[red] (0,0) circle (1.5pt) coordinate (A);
			\fill[red] (3.0ex, 0) circle (1.5pt) coordinate (B);
			\draw[red, thick] (A)--(B);}
	}

\def\bluedimer{\tikz[baseline=-0.5ex]{
			\fill[blue] (0,0) circle (1.5pt) coordinate (A);
			\fill[blue] (3.0ex, 0) circle (1.5pt) coordinate (B);
			\draw[blue, thick] (A)--(B);}
	}

\def\greendimer{\tikz[baseline=-0.5ex]{
			\fill[teal] (0,0) circle (1.5pt) coordinate (A);
			\fill[teal] (3.0ex, 0) circle (1.5pt) coordinate (B);
			\draw[teal, thick] (A)--(B);}
	}

 \def\reddot{\tikz[baseline=-0.5ex]{
			\fill[red] (0,0) circle (1.5pt) coordinate (A);
			}
	}

\def\greendot{\tikz[baseline=-0.5ex]{
			\fill[teal] (0,0) circle (1.5pt) coordinate (A);
			}
	}

\def\bluedot{\tikz[baseline=-0.5ex]{
			\fill[blue] (0,0) circle (1.5pt) coordinate (A);
			}
	}

\def\longstate{\tikz[baseline=-0.5ex]{
			\fill[red] (3.0ex, 0) circle (1.5pt) coordinate (B);
			\fill[teal] (6.0ex,0) circle (1.5pt) coordinate (C);
			\fill[teal] (9.0ex, 0) circle (1.5pt) coordinate (D);
			\fill[red] (12.0ex, 0) circle (1.5pt) coordinate (E);
			\fill[blue] (15.0ex, 0) circle (1.5pt) coordinate (G);
			\fill[teal] (18.0ex, 0) circle (1.5pt) coordinate (H);
			\fill[teal] (21.0ex, 0) circle (1.5pt) coordinate (I);
			\fill[red] (24.0ex, 0) circle (1.5pt) coordinate (J);
			\draw[teal, thick] (C)--(D);
			\draw[red, thick] (B) .. controls (6.0ex, 1.5ex) and (9.0ex, 1.5ex) .. (E);
			\draw[teal, thick] (H) -- (I);
			\node at (3ex, -2.0ex) {$+$};
			\node at (6ex, -2.0ex) {$0$};
			\node at (9ex, -2.0ex) {$0$};
			\node at (12ex, -2.0ex) {$+$};
			\node at (15ex, -2.0ex) {$-$};
			\node at (18ex, -2.0ex) {$0$};
			\node at (21ex, -2.0ex) {$0$};
			\node at (24ex, -2.0ex) {$+$};
	}
}

\begin{document}

\preprint{APS/123-QED}
\title{Hilbert Space Fragmentation in Open Quantum Systems}
\author{Yahui Li}
\affiliation{
Technical University of Munich, 
TUM School of Natural Sciences, 
Physics Department,
Lichtenbergstr. 4,
85748 Garching,
Germany
}%
\author{Pablo Sala}%
\affiliation{Department of Physics and Institute for Quantum Information and Matter, California Institute of Technology, Pasadena, CA 91125, USA}
\affiliation{Walter Burke Institute for Theoretical Physics, California Institute of Technology, Pasadena, CA 91125, USA}

\author{Frank Pollmann}
\affiliation{
Technical University of Munich, 
TUM School of Natural Sciences, 
Physics Department,
Lichtenbergstr. 4,
85748 Garching,
Germany
}%

\date{\today}

\begin{abstract}
We investigate the phenomenon of Hilbert space fragmentation (HSF) in open quantum systems and find that it can stabilize highly entangled steady states.
For concreteness, we consider the Temperley-Lieb model, which exhibits quantum HSF in an entangled basis, and investigate the Lindblad dynamics under two different couplings.
First, we couple the system to a dephasing bath that reduces quantum fragmentation to a classical one with the resulting stationary state being separable.
We observe that despite vanishing quantum correlations, classical correlations develop due to fluctuations of the remaining conserved quantities, which we show can be captured by a classical stochastic circuit evolution.
Second, we use a coupling that preserves the quantum fragmentation structure.
We derive a general expression for the steady state, which has a strong coherent memory of the initial state due to the extensive number of non-commuting conserved quantities.
We show that it is highly entangled as quantified by the logarithmic negativity. 
\end{abstract}

\maketitle


\section{Introduction}

Over the past decades, much effort has been devoted to understanding quantum thermalization in closed systems~\cite{1991_Deutsch, Srednicki_1999, rigol_thermalization_2008, 2016_Alessio}. 
While generic isolated quantum many-body systems are expected to thermalize, a notable counter-example is provided by many-body localization (MBL) --- occurring in  presence of strong disorder~\cite{2006_Basko_MBL, 2007_Oganesyan_Huse_MBL, 2014_Huse_Nandkishore_Oganesyan_MBL, 2015_Nandkishore_MBL, 2019_Abanin_MBL}.
Recently, several alternative mechanisms have been proposed to avoid thermalization in the absence of disorder. 
These include dynamical localization in lattice gauge theories~\cite{2018_Smith_dynamical_localization_LGT, 2018_Brenes_MBL_gauge_invariance}, quantum-many boday scars~\cite{2017_Lukin_atom_scar, 2018_AKLT_scar_sanjay, 2018_scar_exact_excited_states_sanjay, 2021_serbyn_scar}, and Hilbert space fragmentation (HSF) \cite{2020_sala_ergodicity-breaking,2020_khemani_local,2022_Moudgalya_review_HSF_SCAR}. 

The defining property of HSF is the fragmentation of the Hilbert space into \emph{exponentially} many (in system size) dynamically disconnected sectors, known as fragments or Krylov subspaces. 
The non-ergodicity in the case of strong fragmentation, where the largest fragment contains only an exponentially vanishing fraction of the relevant Hilbert space, can lead to infinitely long-lived autocorrelation functions even at infinite temperatures.
More generally, HSF provides a rich playground and leads to distinct physical phenomena, including statistically localized integrals of motion \cite{2020_SLIOMs}, integrable and non-integrable fragments, quantum many-body scars, Krylov-restricted thermalization~\cite{2020_Iadecola_HSF, 2021_scar_fragmentation, Moudgalya_2021_thermalization, 2021_scar_tilted_1d_fermi_hubbard,2022_sanjay_commutant_scar, 2022_Moudgalya_review_HSF_SCAR}, as well as an effective ``Casimir effect''~\cite{2019_HSF_Fractonic_Circuits,2022_Feng_HSF_Casmir} and  universal subdiffusive behavior~\cite{2020_Pablo_automato,  Morningstar_2020, Iaconis_2019, Iaconis_2021,Hart_2022, feldmeier2021critically,Gromov_2020}. The latter has been recently observed in experiments with ultracold atoms~\cite{PhysRevX.10.011042}. 
Moreover, HSF can be used to explain the non-ergodic behavior observed in the experimentally realized tilted systems~\cite{2021_pablo_experiment_fermiHubbard, 2023_pablo_experiment_tilted_Fermi_chain, 2021_morong_SMBL_trapped_ion}, related to Wannier-Stark many-body localization~\cite{2019_Pollmann_SMBL, 2019_Refael_SMBL}. 
Most examples of HSF discussed in the literature exhibit fragmentation in a local product basis (see for example \cite{2020_sala_ergodicity-breaking, 2020_khemani_local, 2020_SLIOMs, 2022_2D_HSF_Nandkishore_Ising,2022_Ryusuke_2D_HSF_weak_tilted_Ising, 2022_Neupert_2D_HSF_subsystem_symmetries, 2023_Feng_HSF_boson_weak_strong_transition}) referred to as \emph{classical fragmentation} (CF). 
In such cases, the fragmented structure can also be realized using classical Markov generators~\cite{Morningstar_2020, Iaconis_2019, 2020_Pablo_automato, Iaconis_2021,Hart_2022, feldmeier2021critically,2022_Lehmann_Pablo_Markov,2022_Frey_HSF_fermihubbard_Markov,2023_Feng_HSF_boson_weak_strong_transition}. 
First examples of fragmentation in an entangled basis --- which we denote \emph{quantum fragmentation} (QF)--- have been only recently proposed~\cite{moudgalya_hilbert_2022}. 
Reference~\cite{moudgalya_hilbert_2022} put forward an algebraic approach using the mathematical notion of \emph{bond} and \emph{commutant} algebras to characterize the set of conserved quantities, which also provides a systematic way to explore the differences among these two types of fragmentation.  

\begin{figure}[bt]
	\includegraphics[width=8cm]{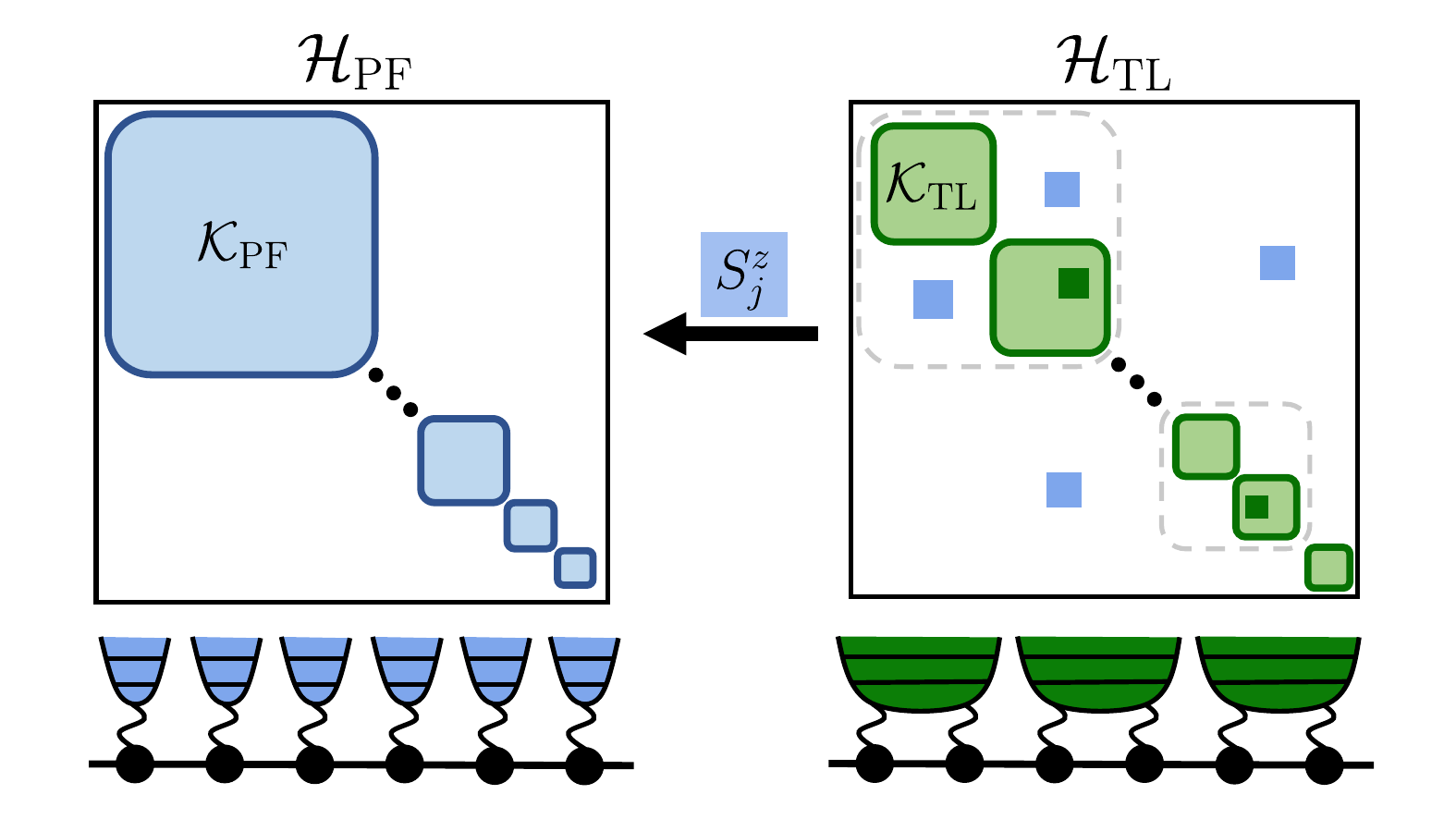}
	\caption{\label{fig:HSF_break}\textbf{Schematic representation of the setup.} The Hilbert spaces of both the pair-flip (PF) and Temperley-Lieb (TL) models fragment into exponentially many Krylov subspaces (solid filled blue and green squares respectively). The degenerate Krylov subspaces of the TL model are contained in the same grey squares. The dephasing noise $L_j = S_j^z$ connects some of the fragmented subspaces of the TL model (on-site dissipative coupling in blue), such that the fragmentation reduces to the classical one of the PF model. Nonetheless, the quantum fragmentation is preserved when using specific two-site dissipative couplings.}
\end{figure}

In realistic settings, quantum many-body systems are never perfectly isolated from their surrounding environment. 
This raises the question to which extent the phenomena related to HSF ---in particular QF, which takes place in an entangled basis---  are affected by couplings to a bath. 
Note that in case of MBL, the localization is destroyed when the system is locally coupled to a dissipative bath~\cite{2016_Dephasing_MBL, 2016_MBL_dissipation, 2016_MBL_dissipation2, 2020_R3_MBL}. 
In Ref.~\cite{2020_essler_open_fragmentation}, CF in open systems due to weak symmetries was studied by exploiting the resulting integrable structure, allowing to obtain the spectrum of the Lindbladian within all invariant subspaces. 
Here we aim to provide an understanding of the generic behavior that quantum fragmented models can display in the presence of a dissipative bath and consider a family of QF models introduced in Ref.~\cite{moudgalya_hilbert_2022}. 
Building up on previous works on the stationary state structure for Lindbladian evolution in the presence of (conventional) conserved quantities \cite{2014_Albert, 2012_Buca_Prosen, 2020_Buca_Zhang}, we investigate systems described by the commutant algebra formalism~\cite{moudgalya_hilbert_2022} and focus on different strong symmetries~\cite{2012_Buca_Prosen, 2020_Buca_Zhang}, i.e., symmetries preserved by both the Hamiltonian and every jump operator. 
We start by considering a dephasing coupling~\cite{book_open_system}, where the system locally couples to a bath that we find to eventually reduce the QF to CF. 
Still, the system preserves a large amount of information of the initial state even at infinite time due to the extensive degeneracy of stationary states.  
On the other hand, a dissipative environment can be engineered and exploited to create exotic non-equilibrium dynamics~\cite{2008_zoller_experiment_quantum_state_pre_open_system, 2008_zoller_Lindblad_pre_state, 2009_verstraete_quantum_state_engineer, 2019_buca_jaksch_non_statioanry_dissipation, 2023_symmetry_induced_DFS}.
For example, Ref.~\cite{2009_verstraete_quantum_state_engineer} proposed to efficiently drive the system to the desired pure state as the unique stationary state by engineering dissipative couplings.
Motivated by this, we consider a fine-tuned coupling which preserves the structure of the QF system. 
Interestingly, we find that the system evolves to a highly entangled stationary state.
Moreover, we propose this as a simple protocol to decide whether a system is quantum fragmented.  

The remainder of the paper is organized as follows. 
In Sec.~\ref{sec:HSF_Lindblad}, we briefly review the commutant and bond algebras formulation for isolated fragmented systems~\cite{moudgalya_hilbert_2022}, and generalize it to open quantum systems focusing on strong symmetries. 
In Sec.~\ref{sec:model}, we then introduce two related fragmented models, the Pair-flip (PF) and the Temperley-Lieb (TL) model, which exhibit classical and quantum fragmentation, respectively. 
We study the TL model under dephasing noise in Sec.~\ref{sec:dephasing}, which leads to a breakdown of quantum fragmentation to the classical fragmentation of the PF model. 
In Sec.~\ref{sec:preserving}, we couple the TL model to the structure-preserving noise, which preserves the original quantum fragmentation of the TL model. 
We analytically derive the stationary states of the dynamics under both couplings that we use to predict saturation values of two-point correlators and two different entanglement measures, the logarithmic negativity, and the operator space entanglement, and compare them with numerical simulations. 
We conclude in Sec.~\ref{sec:conclusion} by summarizing our main findings and discussing open questions. Finally, we consign more technical aspects of our work to the appendices.

\section{Methods: Commutant algebras and Lindblad dynamics}\label{sec:HSF_Lindblad}
In this section, we introduce the methods used to investigate the role of HSF in open quantum many-body systems.
First, we review the mathematical framework introduced in Ref.~\cite{moudgalya_hilbert_2022} that characterizes HSF in closed systems in terms of \emph{bond and commutant algebras}.
Second, we discuss the role of symmetries in the context of Lindblad dynamics generalizing the commutant algebra formulation.

\subsection{Commutant and bond algebras}
The phenomenon of HSF arises as a consequence of certain constraints being imposed on the dynamics of many-body systems.
Given a family of Hamiltonians $H = \sum_j J_j h_j$ parameterized by real coefficients $\{J_j\}$, fragmentation is a property that is completely characterized by the local terms $\{h_j\}$ and thus holds for any choice of coefficients.
This distinguishes HSF from other symmetries that might appear for certain choices of $J_j$, such as translation invariance with uniform $J_j$. 
Reference~\cite{moudgalya_hilbert_2022} formalized this observation using the language of \emph{bond} and \emph{commutant} algebras for isolated quantum systems, which we will review in the following. 
A \emph{bond algebra} $\mathcal{A}$ is the algebra generated by arbitrary linear combinations of products of the local terms $\{h_j\}$, together with the identity operator $\mathbb{1}$. 
The corresponding \emph{commutant algebra} $\mathcal{C}$ is the set of conserved quantities, namely the centralizer of $\mathcal{A}$ including all operators that commute with every local term
\begin{equation}
\mathcal{A} := \langle \{h_j\} \rangle, \ 
\mathcal{C} := \{O:\, [O, h_j] = 0, \forall j\}.
\end{equation}
We refer to the latter using the shorthand notation $\mathcal{C}=\langle \{h_j\} \rangle'$.
Both $\mathcal{A}$ and $\mathcal{C}$ are von Neumann algebras, i.e., they include the identity operator and are closed under conjugation~\cite{1998_lecture_von_Neumann_algebra}.
Importantly, every element in $\mathcal{C}$ commutes with every element in $\mathcal{A}$, i.e., they are the centralizers of each other. 
As such the Hilbert space can be decomposed into irreducible representations of $\mathcal{C}\times \mathcal{A}$~\cite{1998_lecture_von_Neumann_algebra, 2017_math_von_Neumann}, 
\begin{equation}\label{eq:Hilbert}
\mathcal{H} = \bigoplus_{\lambda} \left( \mathcal{H}_\lambda^{(\mathcal{C})} \otimes \mathcal{H}_\lambda^{(\mathcal{A})}  \right), 
\end{equation}
where $\mathcal{H}_\lambda^{(\mathcal{C})}$ and $\mathcal{H}_\lambda^{(\mathcal{A})}$ are the $d_\lambda$ and $D_\lambda$ dimensional irreducible representations of $\mathcal{C}$ and $\mathcal{A}$, respectively. 
This decomposition implies that the elements of the bond algebra $h_\mathcal{A}\in\mathcal{A}$, generate independent dynamics within $\mathcal{H}_\lambda^{(\mathcal{A})}$ while acting trivially on $\mathcal{H}_\lambda^{(\mathcal{C})}$.
Therefore, for fixed $\lambda$, there are $d_\lambda$ \emph{degenerate} Krylov subspaces or fragments with dimension $D_\lambda$. 
We will denote the degenerate Krylov subspaces as $\mathcal{K}_\alpha^\lambda$, with $\alpha = 1,...,d_\lambda$, and omit $\lambda$ if there is no degeneracy.

The formulation in terms of bond and commutant algebras provides a unifying framework to describe the decomposition of the Hilbert space, which applies to both conventional and unconventional symmetries like HSF~\cite{moudgalya_hilbert_2022, 2022_sanjay_commutant_symmetries}. 
The difference appears in the  scaling of the dimension of the commutant $\dim(\mathcal{C}) = \sum_\lambda d_\lambda^2 $ with system size: It scales exponentially for fragmented systems while at most polynomially for conventional symmetries.
When the commutant $\mathcal{C}$ is Abelian, every irreducible representation is one-dimensional ($d_{\lambda} = 1 $) and hence, the Hilbert space reduces to a direct sum of non-degenerate Krylov subspaces, $\mathcal{H} = \bigoplus_{\alpha} \mathcal{K}_{\alpha}$.  
Projectors $\Pi_{\alpha} = \sum_{\beta} |\psi_{\alpha\beta}\rangle \langle \psi_{\alpha \beta}|$ onto those subspaces span the commutant, where $\{| \psi_{\alpha\beta} \rangle\}$ is an orthonormal basis in $\mathcal{K}_{\alpha}$. 
On the other hand, non-Abelian commutants include larger dimensional irreducible representations  $d_\lambda>1$, corresponding to degenerate Krylov subspaces. 
In this case, the projectors $\Pi_{\alpha}^{\lambda}$ onto different Krylov subspaces $\mathcal{K}_{\alpha}^{\lambda}$ span a maximal Abelian subalgebra of $\mathcal{C}$, while the full $\mathcal{C}$ is generated by not only the projectors but also the intertwine operators between degenerate ones, $\Pi_{\alpha \alpha^\prime}^{\lambda} = \sum_{\beta} |\psi_{\alpha \beta}^\lambda \rangle \langle \psi_{\alpha^\prime \beta}^\lambda|$~\cite{moudgalya_hilbert_2022}.
For example, the commutant algebra of SU$(2)$-symmetric systems is non-Abelian and contains non-commuting conserved quantities such as $S^x_{\text{tot}}$, $S^y_{\text{tot}}$, and $S^z_{\text{tot}}$. 
The total spin representation $\lambda$ is given by the eigenvalues of $(\vec{S}_{\textrm{tot}})^2$ as $\lambda(\lambda+1)$.
There are $d_\lambda = 2\lambda + 1$ degenerate Krylov subspaces with the same $\lambda$, which are labeled by different spin-$z$ projections $S^z_{\text{tot}} = -\lambda, -\lambda+1, ... \lambda$, leading to the Hilbert space decomposition as in Eq.~\eqref{eq:Hilbert}~\cite{2007_Bartlett_SU2_decompose_Hilbert_space}.

Fragmentation can be classified as either classical or quantum. 
A system is said to be classically fragmented if one can find a common eigenbasis of product states for all elements in a maximal Abelian subalgebra of the commutant. 
This means that the Krylov subspaces can be spanned by a product state basis.
Otherwise a system is said to be quantum fragmented. 
By this definition, we associate CF with the existence of a basis of product states and QF with an entangled basis---which is different from the commutant being Abelian or not. 
Specifically, an entangled basis can also appear for Abelian commutants. 
For example, for an SU$(2)$-symmetric system, adding the term $S^z_{\mathrm{tot}}$ preserves the Hilbert space structure in an entangled basis but breaks the degeneracy of the Krylov subspaces, leading to the so-called dynamical SU(2) symmetry and an Abelian commutant~\cite{2022_sanjay_commutant_symmetries}. 
Note that the current definition of QF is still not ideal since it leaves some room for ambiguous or trivial examples.

\subsection{Lindblad dynamics of fragmented systems}
We study the dynamics of fragmented systems coupled to a Markovian bath described by a Lindblad master equation, $\frac{d \rho}{dt} = \mathcal{L}(\rho)$ (see Fig.~\ref{fig:HSF_break}). Here $\mathcal{L}$ is a Liouvillian superoperator with \cite{book_open_system, 2020_intro_Lindblad}
\begin{equation}\label{eq:Lindblad_state}
    \mathcal{L}(\rho) = -i[H,\rho] + \sum_j \gamma _j \left(L_j\rho L_j^\dagger -\frac{1}{2} \{L_j^\dagger L_j,\rho\}\right),
\end{equation}
where the positive coefficients $\gamma_j$ correspond to the decay rates, $\{L_j\}$ are jump operators describing the coupling to a bath, and we set $\hbar=1$.
Equivalently, the time evolution of an operator in the Heisenberg picture is generated by the adjoint of the Liouvillian superoperator, $\frac{d O}{dt} = \mathcal{L}^\dagger (O)$.

Of particular interest to us is the stationary state
\begin{equation} \label{eq:rho_ss_1}
    \rho_{\text{ss}} = \lim_{t\rightarrow\infty} e^{t\mathcal{L}} \rho_0 
\end{equation}
as an eigenstate of $\mathcal{L}$ with zero eigenvalue, namely $\mathcal{L}(\rho_{\text{ss}})= 0$. 
Generally, for a Liouvillian without symmetries, there is a unique stationary state and it preserves no information of the initial state.
On the other hand, symmetries and conserved quantities can lead to multiple stationary states and a memory effect in the long time limit~\cite{2012_Buca_Prosen, 2014_Albert, 2016_Albert_geometry_and_response_of_Lindbladians, 2020_Buca_Zhang}. 
A simple case is the presence of a \emph{strong} unitary symmetry $S$ that is preserved by both the Hamiltonian and every jump operator, i.e., $[S, H] = [S, L_j] = [S,L_j^\dagger]=0$ for all $j$~\cite{2012_Buca_Prosen, 2020_Buca_Zhang}. 
The space of bounded operators $\mathcal{B}(\mathcal{H})$ decomposes into orthonormal subspaces, $\mathcal{B}_{\alpha\alpha^\prime} = \text{span}\{|\psi_{\alpha}\rangle \langle \psi_{\alpha^\prime}|\}$, where $|\psi_{\alpha}\rangle$ is an eigenstate of $S$ with eigenvalue $s_\alpha$. 
Each subspace labeled by different quantum numbers of $S$ evolves independently since $\mathcal{L} \mathcal{B}_{\alpha\alpha^\prime} \subseteq \mathcal{B}_{\alpha\alpha^\prime}$. 
Thus, the stationary state inherits the block diagonal structure given by the symmetry, which leads to at least as many distinct stationary states as the number of symmetry sectors~\cite{2012_Buca_Prosen, 2020_Buca_Zhang}.

Let us now investigate the phenomenon of HSF in the strong symmetry sense.
In Lindblad systems, the dynamics is generated by the Hamiltonian $H=\sum_j J_j h_j$ and the jump operators $\{L_j\}$. Reference~\cite{Baumgartner_2008_2} considered the commutant $\langle H, \{L_j\}, \{L_j^\dagger\}\rangle^\prime $ associated with the (total) Hamiltonian and the jump operators, which was shown to give a complete set of conserved projectors onto mutually orthogonal subspaces with independent dynamics in $\mathcal{B}(\mathcal{H})$.
Note however that the analysis only applies when the conserved operators form an algebra\footnote{See counterexamples of conversed operators not forming an algebra in Ref.~\cite{Baumgartner_2008_2}.}.
In the language of Ref.~\cite{2022_sanjay_commutant_symmetries}, $\langle H, \{L_j\}, \{L_j^\dagger\}\rangle$ corresponds to a \emph{local} algebra rather than to a bond algebra, since $H$ is an extensive sum of local terms.  
To extend the analysis of HSF in terms of bond and commutant algebras to open quantum systems, we define the \emph{open bond algebra} $\mathcal{A}^O=\langle \{h_j\}, \{L_j\}\rangle$ where we focus on Hermitian $L_j$, and the corresponding open commutant $\mathcal{C}^{O} = \langle\{h_j\}, \{L_j\}\rangle'$.
Hence, the conserved projectors $\{\Pi^\lambda_{\alpha}\} \in \mathcal{C}^O$ then satisfy $[\Pi^\lambda_{\alpha}, h_j] = [\Pi^\lambda_{\alpha}, L_j] = 0$ for all $j$, such that each subspace evolves independently.
Note that these $\Pi^\lambda_\alpha$ project onto minimal (irreducible) subspaces of the dynamics generated by $\mathcal{L}$, as they span the maximal Abelian subalgebra of the open commutant~\cite{Baumgartner_2008_2}.
All together we find that the operator space $\mathcal{B}(\mathcal{H})$ decomposes into orthogonal, invariant, minimal subspaces $\mathcal{B}_{\alpha\alpha^\prime}$~\cite{Baumgartner_2008, Baumgartner_2008_2}. 
As stated above, the existence of non-unique stationary states is now guaranteed by these subspaces, where now the degeneracy of the stationary state scales exponentially with system size due to HSF.

\section{Model and setup}\label{sec:model}
We study the dynamics of quantum fragmented systems coupled to a dissipative environment by considering the family of Temperley-Lieb (TL) models as a concrete example.

First, we introduce the closely-related spin-$1$ pair-flip (PF) model, which exhibits CF, i.e., it is fragmented in a product-state basis.
The Hamiltonian is given by  
\begin{equation}\label{eq:H_PF}
\begin{split}
    H_{\text{PF}} =& \sum_{j=1}^{N-1} \sum_{\alpha,\beta \in \{+,0,-\}} [g_{j,j+1}^{\alpha\beta} (|\alpha\alpha\rangle \langle \beta\beta|)_{j,j+1} + \text{h.c.}] \\ &+ \sum_{j=1}^N\sum_{\alpha\in\{+,0,-\}}l_{j\alpha} (|\alpha\rangle\langle\alpha|)_j,  
\end{split}
\end{equation}
where $\alpha$, $\beta$ denote different spin-$z$ components $\{-,0,+\}$, and $g_{j,j+1}^{\alpha\beta}$ and $l_{j\alpha}$ are arbitrary real coefficients. We assume  open boundary conditions (OBC) and even number of sites for convenience.
The constrained dynamics of the PF model can be visualized by mapping  product states in the computational basis to colored pairs and dots.
Specifically, we denote the spins with different colors as $|\reddot\rangle = |+\rangle$, $|\greendot\rangle = |0\rangle$, and $|\bluedot\rangle = |-\rangle$. 
Using this representation, the pair-flip terms of $H_\text{PF}$ change neighboring spins with the same color, 
\begin{equation}
|\reddimer\rangle \leftrightarrow  |\greendimer\rangle \leftrightarrow |\bluedimer\rangle.
\end{equation}
The PF model has two independent U$(1)$ charges, which are given by $N^+ = \sum_j (-1)^j N_j^+$ and $N^- = \sum_j (-1)^j N_j^- $, with $N_j^\alpha = (|\alpha\rangle\langle \alpha|)_j$. 
These U$(1)$ symmetry sectors further split into smaller Krylov subspaces labeled by a non-local invariant~\cite{2018_PFmodel}, which we will discuss in the following.
Starting from a product state, we first connect all the adjacent spins with the same color from left to right. 
Next we remove the paired spins and repeat the first step until there are only unpaired spins with a different color from their nearest neighbors to the left. 
The unpaired spins are then referred to as \emph{dots}. 
We denote dot patterns of size $2\lambda$ as $A_{\lambda}$. 
Let us for example consider a state with the dot pattern $(\bluedot\,\,\reddot)$, 
\begin{equation}
|\longstate\rangle.\label{dotpattern}
\end{equation}
We observe that dot patterns, i.e., the color and sequence of unpaired spins, are invariant under the action of a pair-flip, providing non-local and mutually commuting conserved quantities similarly to Ref.~\cite{2020_SLIOMs}. 
Thus, each Krylov subspace can be labeled by a dot pattern. 
The number of different dot patterns grows exponentially with system size and thus the Hilbert space fragments into exponentially many Krylov subspaces in the local $z$ basis.
Since the fragmentation occurs in a basis of product states ---common eigenbasis of all elements of the Abelian commutant--- the PF model exhibits CF. See schematic representation appearing in Fig.~\ref{fig:HSF_break}.

Next we introduce the SU$(3)$  symmetric spin-$1$ TL model \cite{1990_TL, 2010_TL}, which is a special case of the PF model where all coupling strengths for different color pairs are the same.
The Hamiltonian is given by~\footnote{Note that the TL model can be mapped to the purely-biquadratic model $H = \sum_j J_j (\vec{S}_j \cdot \vec{S}_{j+1})^2$ by a local unitary operator, $U=\prod_{j \text{ odd}} \exp{(i\pi S_j^y)}$~\cite{moudgalya_hilbert_2022}.}
\begin{equation}
    H_{\text{TL}} =  \sum_j J_j e_{j,j+1},
\end{equation}
where $e_{j,j+1}=\sum_{\alpha,\beta \in \{+,0,-\}} (|\alpha\alpha\rangle\langle\beta\beta|)_{j,j+1}$ project onto the dimer state $|\dimer\rangle \equiv \frac{1}{\sqrt{3}}|++\rangle + |00 \rangle + |--\rangle$. These local terms $\{e_{j,j+1}\}$ generate the bond algebra $\mathcal{A}_{\text{TL}}$, which is the so-called Temperley-Lieb algebra~\cite{1990_TL, 2010_TL}.

As a special case of the PF model, the TL model is at least as (classically) fragmented as the PF model. 
In addition to the SU$(3)$ symmetry, the constrained dynamics conserves extended dot patterns, including colored dot patterns of the PF model, e.g., $|\bluedot\,\,\,\,\,\reddot\rangle_{j,k}$, and additional entangled dot patterns, e.g.,  $|\entangleddota\rangle_{j,k} = \frac{1}{\sqrt{2}}(|++\rangle-|--\rangle)_{j,k}$~\cite{moudgalya_hilbert_2022}. 
Note that the choice of the dot states is not unique due to the non-Abelian nature of $\mathcal{C}$. 
The following example shows that the dot pattern $|\entangleddota\rangle_{j,k}$ is conserved: 
\begin{equation}
    e_{j,j+1} |\onedimera\rangle = |\entangleddimerdot\rangle.
\end{equation}
The TL model is then block-diagonal in an entangled basis given by the dimers and dots configurations (more examples are shown in App.~\ref{subapp:TL_basis}).
Thus the TL model exhibits QF.
Moreover, the resulting commutant algebra $\mathcal{C}_{\text{TL}}$ is non-Abelian and the dimension of the irreducible subspaces are $d_\lambda \geq 1$~\cite{2010_TL, moudgalya_hilbert_2022}. 
Therefore, there are $d_\lambda$ degenerate Krylov subspaces for fixed $\lambda$, which are labeled by different dot patterns with the same $2\lambda$ length.
Note that in the previous discussion, we distinguished between CF and Abelian commutant, as well as between QF and non-Abelian commutant. For example, one can find systems with an Abelian commutant which nonetheless require an entangled basis~\cite{2022_sanjay_commutant_symmetries}. 

Following the distinction between strong and weak fragmentation as discussed in Ref.~\cite{2020_sala_ergodicity-breaking}, we verify that both the PF model and the TL model exhibit strong fragmentation with respect to the full Hilbert space. The dimension of the largest Krylov subspace scale as $D_{\text{max}}/3^N\sim \exp (-aN)$ with $a<1$. See App.~\ref{subapp:fss_ac} for additional details.

\begin{figure}[bt]
	\centering
	\includegraphics[width=8.0cm]{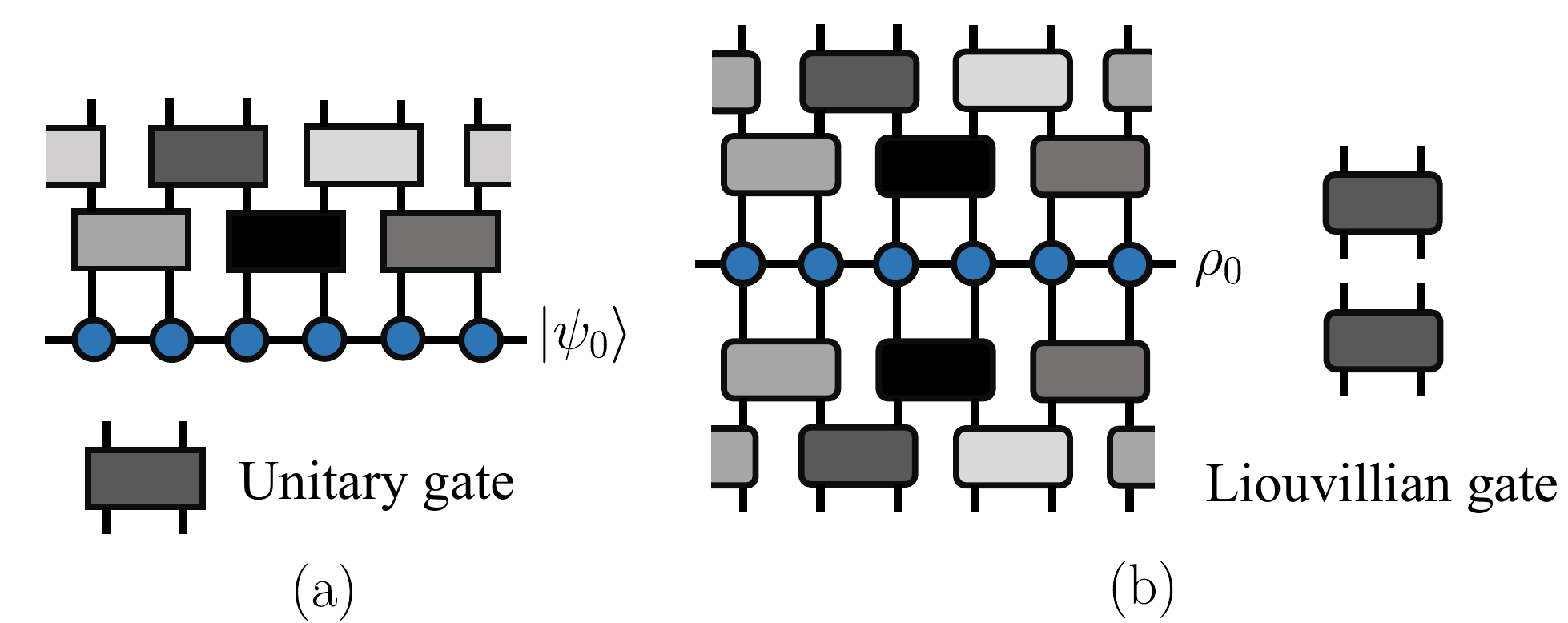}
	\caption{\textbf{Lindblad random circuits.} A single time step for random circuit evolution for (a) a closed system with two-site unitary gates $U_{\tau, j} = e^{-iJ_j h_{j,j+1}}$; and (b) the Lindblad evolution with Liouvillian gates $\mathcal{U}_{\tau, j} = e^{\mathcal{L}_{j,j+1}}$, where $\{J_j\}$ are random coefficients extracted from a uniform distribution $J_j \in [0.8, 1.2]$. Here, the blue circles represent the initial state or density matrix.}
	\label{fig:Lindblad_simulation}
\end{figure}

To study the effect of fragmentation on the Lindblad evolution, we discretize the dynamics and implement a local random quantum circuit including both Hamiltonian and Lindblad evolutions. This implementation breaks energy conservation and translation symmetry, and only preserves those quantities belonging to the commutant algebra. The setting of random circuits is shown in Fig.~\ref{fig:Lindblad_simulation}. Every time step includes two consecutive layers of non-overlaping gates, such that after $t$ time steps the time evolution is given by $\mathcal{U}(t; 0) = \prod_{\tau=1}^{t} \mathcal{U}_{\tau}$ with
\begin{align}\label{eq:Lindblad_circuit}
		 \mathcal{U}_{\tau}	=
			\prod_{j\ \text{even}}\mathcal{U}_{\tau,j}  
			\prod_{j\ \text{odd}}\mathcal{U}_{\tau,j}.  
\end{align}
The Liouvillian gates are superoperators given by $\mathcal{U}_{\tau,j} = e^{\mathcal{L}_{j,j+1}}$ (see Fig.~\ref{fig:Lindblad_simulation}b), with
\begin{equation}
    \mathcal{L}_{j,j+1}(\rho) = -iJ_j[h_{j,j+1},\rho] + \mathcal{D}_{j,j+1}(\rho),
\end{equation}
where $\{J_j\}$ are uniformly distributed random coefficients for different sites $j$ and time steps $\tau$. The dissipation term is $\mathcal{D}_{j, j+1}(\rho) = \gamma \sum_{l=j,j+1} (L_l\rho L_l^\dagger -\frac{1}{2} \{L_l^\dagger L_l,\rho\})$ for one-site jump operators, and $\mathcal{D}_{j, j+1}(\rho) = \gamma (L_{j, j+1}\rho L_{j, j+1}^\dagger -\frac{1}{2} \{L_{j, j+1}^\dagger L_{j, j+1},\rho\})$ for two-site jump operators. In the following, we use the dimensionless hoppings $J_j$ to be uniformly distributed in the interval $[0.8, 1.2]$. Moreover, we choose $\gamma_j=\gamma$ as this does not affect our results~\cite{Baumgartner_2008}. When $\gamma=0$, the Liouvillian gates become random unitaries with the overall phase fluctuating around $\pi$.
This implementation allows us to compare our numerical results with the analytic prediction obtained using the formalism introduced in the previous section. 

\section{Dephasing noise}\label{sec:dephasing}
We first consider a dephasing noise given by $L_j = S_j^z$. For many-body localized systems, such coupling delocalizes the system and drives it to an infinite temperature state $\rho \propto \mathbb{1}$~\cite{2016_MBL_dissipation, 2016_MBL_dissipation2, 2016_Dephasing_MBL, 2020_R3_MBL}. For the TL model, however, the dephasing noise preserves the CF while breaking the QF. When considering the whole Hilbert space, this turns into non-ergodic behavior and extensively degenerate stationary states. 

The mechanism for the breakdown from the QF of the TL model to the CF of the PF model is shown in Fig.~\ref{fig:HSF_break}. Intuitively, the TL model is symmetric with respect to different color pairs due to the SU$(3)$ symmetry, while the dephasing noise distinguishes different colors. However, this respects the CF, as the jump operators are elements of the PF bond algebra, $S_j^z \in \mathcal{A}_{\text{PF}}$. Moreover, any element of this algebra can be written as linear combinations of products of elements in the TL bond algebra \emph{and} the dephasing jump operators as explicitly shown in App.~\ref{subapp:vonNeumann_deph}. Therefore, the corresponding bond algebra is given by the PF one $\langle\{h_j\}, \{S_j^z\}\rangle = \mathcal{A}_{\text{PF}}$ with open commutant  $\mathcal{C}^O = \mathcal{C}_{\text{PF}}$. This implies that the symmetries of the Liouvillian are those of the PF model. In this section, we study the effect of the breakdown of quantum fragmentation, and sketch the derivation of the stationary state for this case.

\subsection{Stationary states with classical fragmentation}
We now derive the stationary state of the TL model under dephasing noise. 
As we just showed, both the Hamiltonian and the jump operators preserve the CF of the PF model, with $[h_j, \Pi_{\alpha}] = [L_j, \Pi_{\alpha}] = 0, \forall j$.
Therefore, the operator space is decomposed into orthogonal subspaces with independent dynamics,
\begin{equation}
    \mathcal{L}(\Pi_{\alpha}\rho\Pi_{\alpha^\prime}) = \Pi_{\alpha}\mathcal{L}(\rho) \Pi_{\alpha^\prime},
\end{equation}
or equivalently, $\mathcal{L}\mathcal{B}_{\alpha \alpha^\prime}\subseteq \mathcal{B}_{\alpha \alpha^\prime}$, where we denote the diagonal subspaces $\mathcal{B}_{\alpha} \equiv \mathcal{B}_{\alpha\alpha}$. This is the natural extension of strong symmetry for fragmented systems. 

Next, we show that there is a unique stationary state within each $\mathcal{B}_{\alpha}$. 
As all jump operators are Hermitian, the infinite temperature state $\rho \propto \mathbb{1}$ is a stationary state in $\mathcal{B}(\mathcal{H})$. 
Therefore, there exists a stationary state $\mathbb{1}_\alpha/D_{\alpha}\equiv\Pi_{\alpha}\mathbb{1}/D_{\alpha}$ within each $\mathcal{B}_{\alpha}$, with $\mathcal{L}(\mathbb{1}_\alpha/D_{\alpha})=0$. 
This is because the dissipation induces full decoherence within each invariant subspace. 
Moreover, as the projectors $\{\Pi_{\alpha}\}$ span the maximal Abelian subalgebra of the open commutant~\cite{moudgalya_hilbert_2022}, these invariant subspaces $\mathcal{B}_{\alpha}$ are minimal subspaces~\cite{Baumgartner_2008_2}. 
Therefore, the stationary state within each $\mathcal{B}_{\alpha}$ is unique~\cite{Baumgartner_2008_2}. Additional details can be found in App.~\ref{subapp:stationary_states}.

Combining the stationary state structure (i.e., a unique stationary state within each minimal subspace) and the corresponding conserved quantities $\{\Pi_{\alpha}\}$, we find that the general expression of the stationary state is given by
\begin{equation}\label{eq:rho_ss_classical}
    \rho_{\text{ss}} = \bigoplus_{\alpha} c_\alpha \frac{\mathbb{1}_\alpha}{D_\alpha}, c_\alpha = \mathrm{Tr}(\Pi_{\alpha} \rho_0).
\end{equation}
The coefficients $c_\alpha\in \mathbb{R}$ are the weights of the initial state within the diagonal subspaces $\mathcal{K}_\alpha$. 
The stationary state preserves the weight $c_\alpha$, while all the off-diagonal (coherent) information is lost. A more detailed derivation can be found in App.~\ref{subapp:stationary_states}. Due to fragmentation, the number of distinct stationary states reached by different initial states scales exponentially with the system size, signalling a strong memory effect. In the following, using the expression in Eq.~\eqref{eq:rho_ss_classical}, we analyze the long-time behavior of the TL model under dephasing noise.

\begin{figure}[t]
\includegraphics[width=8cm, scale=1.0]{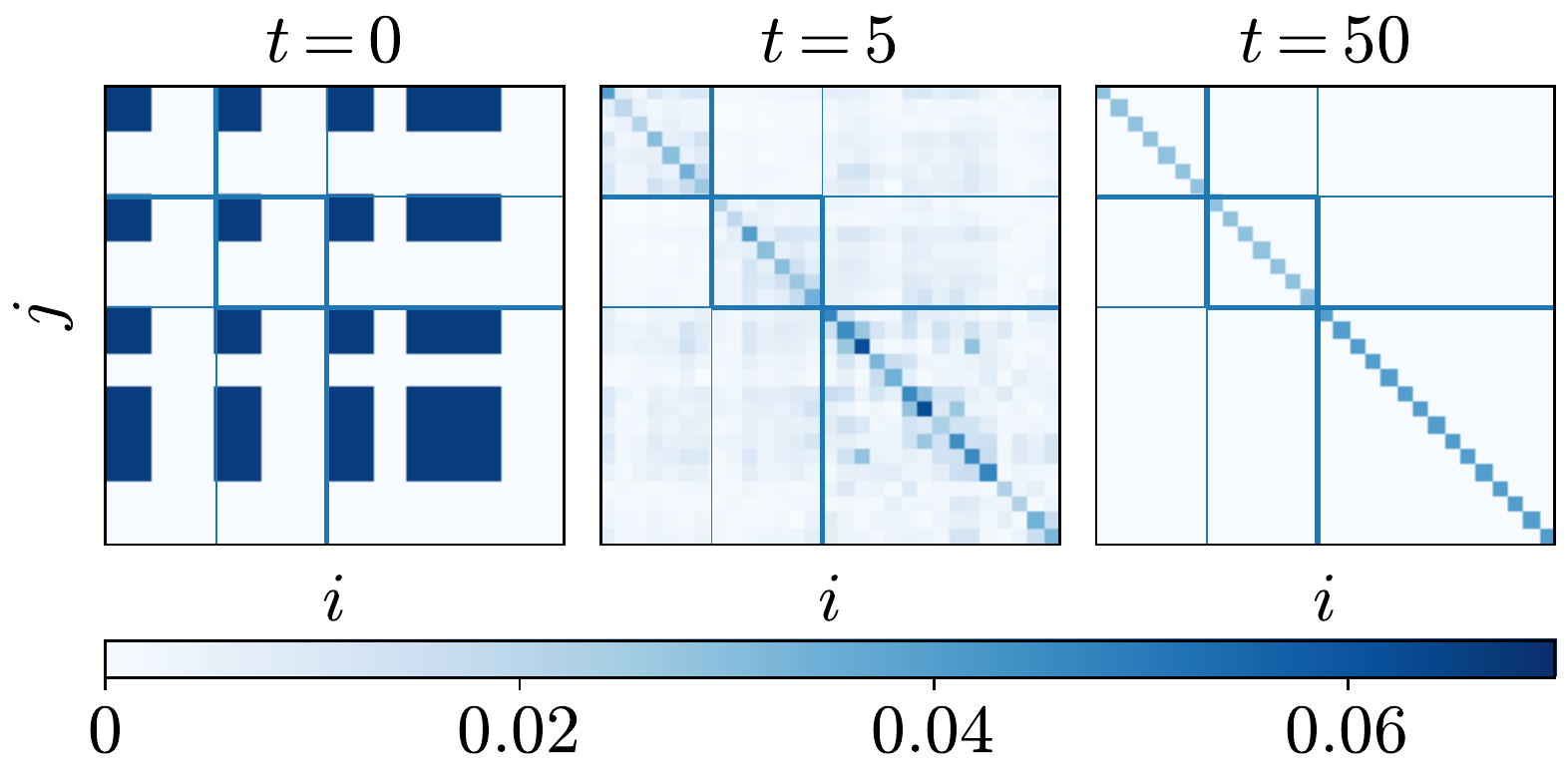}
\caption{\label{fig:TL_rho_Zj} \textbf{Time evolution of the density matrix under dephasing noise.} Time evolution of the density operator under dephasing noise $L_j = S_j^z$ using ED, with system size $N=4$ and $\gamma=1$. The initial state is specified in Eq.~\eqref{eq:initial_state_two_degenerate}, having nonzero overlap with three different Krylov subspaces. The color intensity in the figures is the magnitude of matrix elements $|\rho_{ij}|$ in the product $z$ basis, and the solid blue lines separate different Krylov subspaces. }
\end{figure}

Figure~\ref{fig:TL_rho_Zj} shows an example of the evolution of the density operator (written in the local $z$-basis) by exact diagonalization (ED). To compare with the case of quantum fragmentation we use the initial state 
\begin{equation}\label{eq:initial_state_two_degenerate}
    |\psi_0\rangle = \frac{1}{\sqrt{3}}\left(|\dimer \ \dimer \rangle + |\dimer\rangle |\twodota\rangle + |\dimer\rangle |\twodotb\rangle\right),
\end{equation}
which has non-zero overlap only with three Krylov subspaces: the fully-paired subspace (with zero dots) and other two labeled by the dot patterns $(\bluedot \ \reddot)$ and $(\reddot \ \bluedot)$. At long times, all off-diagonal matrix elements vanish. The stationary state is then the direct sum of projected identities within the diagonal blocks, with the weight determined by the initial state.

\subsection{Infinite temperature autocorrelation function}
In this section, we investigate the effect of fragmentation on infinite-temperature autocorrelation functions 
\begin{equation}
     \langle O(t) O(0) \rangle_c \equiv \langle O(t) O(0) \rangle - \langle O(t) \rangle \langle O(0)\rangle,
\end{equation}
under Lindblad evolution, where $\langle O \rangle \equiv \mathrm{Tr}(\rho O)$. The evolution of an operator $O$ is given by $O(t) = e^{t\mathcal{L}^\dagger} (O)$, which reduces to $O(t) = e^{iHt} O e^{-iHt}$ without dissipation. For the observables we consider in the following,  the disconnected part is always zero.

For closed systems, the infinite-time average of autocorrelation functions is lower bounded by the Mazur bound~\cite{MAZUR1969, SUZUKI1971, Mazur_2021}, which relates a finite saturation value with the presence of conserved quantities. For example, for the family of PF models and considering the local observable $O=S_j^z$ in a closed system, this bound is given by~\cite{moudgalya_hilbert_2022}
\begin{equation}\label{eq:Mazur_classical}
    M_{\text{PF}}(S_j^z) = \frac{1}{3^N} \sum_\alpha \frac{[\mathrm{Tr}(\Pi_\alpha S_j^z)]^2}{D_{\alpha}},
\end{equation}
where $D_{\alpha}$ is the dimension of Krylov subspace $\mathcal{K}_\alpha$. Here $\{\Pi_{\alpha}\}$ span a full set of conserved quantities for the Abelian commutant $\mathcal{C}_{\text{PF}}$. Ref.~\cite{moudgalya_hilbert_2022} numerically found that the bound $M_{\text{PF}}$ scales as $1/N$ in the bulk, hence vanishing in the thermodynamic limit.

\begin{figure}[t]
\includegraphics[width=8cm]{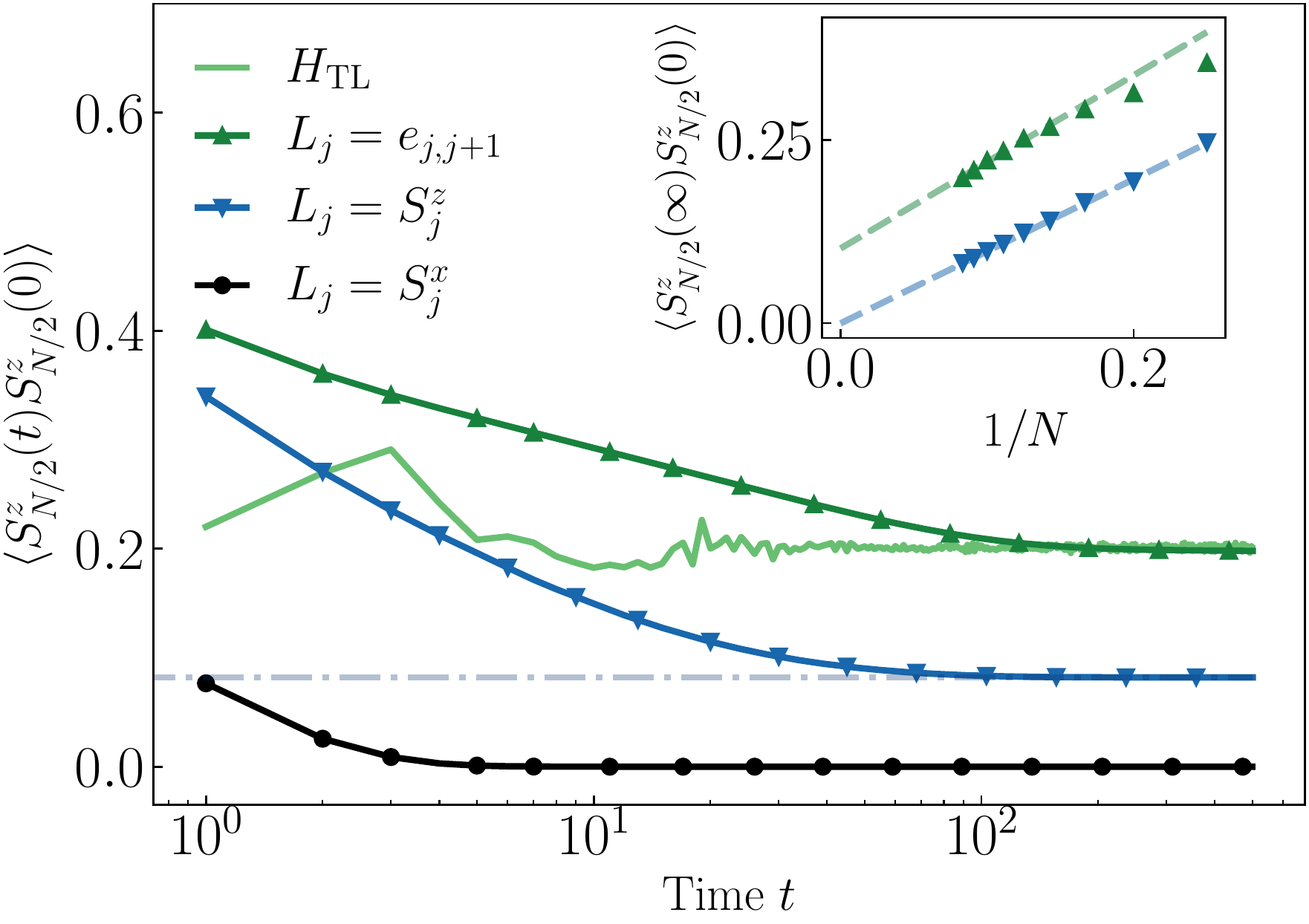}
\caption{\label{fig:Lindblad_autocorrelation} \textbf{Infinite-temperature autocorrelation functions.} Time evolution of bulk autocorrelation functions $\langle S_{N/2}^z(t) S_{N/2}^z(0)\rangle$ for unitary and open quantum dynamics using different jump operators $L_j$. We use a system size $N=12$ and $\gamma=1$. The unitary dynamics is calculated using ED, while we use TEBD for the Lindblad dynamics with bond dimension $\chi=256$. Under the noise $L_j = e_{j,j+1}$ that preserves the quantum fragmentation, the autocorrelation function saturates to the same value as in the closed system for the TL model $H_{\text{TL}}$. Under the dephasing noise $L_j = S_j^z$, the autocorrelation function saturates to a finite value, which is the Mazur bound of the PF model (blue dotted-dashed line). The spin-flip noise $L_j = S_j^x$ further destroys the classical fragmentation, which leads to vanishing autocorrelation functions.}
\end{figure}

In Fig.~\ref{fig:Lindblad_autocorrelation}, we show the evolution of infinite-temperature autocorrelation functions $\langle S_{N/2}^z(t) S_{N/2}^z(0) \rangle$ of the TL model for both closed and open quantum dynamics under different dissipative couplings. 
For closed systems (green solid line), we numerically evaluate the infinite-temperature correlations by uniformly sampling initial Haar random states as prescribed by quantum typicality \cite{2007_Typicality, 2015_Typicality}, which saturates to a finite value. 
For the open dynamics, we simulate the Lindblad evolution using the time-evolving block decimation (TEBD) algorithm~\cite{2003_TEBD, 2004_TEBD_MPO, 2004_TEBD_MPO_Vidal}, with the infinite temperature configuration as initial state $\rho_0 \propto \mathbb{1}$. 
Under dephasing noise (down-pointing triangles), we find that the autocorrelation function saturates to a lower value than the TL model in closed systems, indicating that the dephasing noise reduces the symmetries of the TL model. 
The saturation value is exactly the Mazur bound $M_\text{PF}$ in closed systems given by Eq.~\eqref{eq:Mazur_classical} (blue doted-dashed line). 
In the inset of Fig.~\ref{fig:Lindblad_autocorrelation}, we numerically verify that the saturation values decay as $1/N$ as previously found in Ref.~\cite{moudgalya_hilbert_2022}. 
Appendix~\ref{subapp:fss_ac} contains additional results for boundary correlations, where a finite saturation value is found.

This agreement between the saturation of autocorrelation of the TL model under dephasing noise and the PF Mazur bound can be explained using the same analysis as for the stationary state $\rho_{\textrm{ss}}$ but now for the stationary value of an operator $O(\infty) = \lim_{t\rightarrow\infty} e^{t\mathcal{L}^\dagger} O$, which is given by
\begin{equation}
    O(\infty) = \bigoplus_\alpha O_\alpha \frac{\mathbb{1}_{\alpha}}{D_\alpha}, \ O_\alpha = \mathrm{Tr}(\Pi_{\alpha}O).
\end{equation}
Here $O_\alpha$ is a constant given by the overlap of the operator $O$ and the projector. Using $O = S_j^z$, we obtain the saturation value of $\langle S_j^z(\infty) S_j^z(0) \rangle_c$ as the inner product between $S_j^z$ and its stationary value $S_j^z(\infty)$ recovering Eq.~\eqref{eq:Mazur_classical}. This explains why the autocorrelation function under dephasing noise saturates exactly to the Mazur bound for the PF model. We provide a different proof to the same result in App.~\ref{subapp:Mazurbound_open}, by generalizing the Mazur bound to open systems for diagonalizable $\mathcal{L}$ with strong symmetries.

\subsection{Logarithmic negativity}
We now investigate the spreading of quantum correlations across the system using the logarithmic negativity~\cite{2005lognegativity}, an entanglement measure for mixed states defined as
\begin{equation} \label{eq:negat}
    E_{\mathcal{N}} = \log \|\rho^{T_B}\|_1.
\end{equation}
Here $\|A\|_1 = \mathrm{Tr}\sqrt{A^\dagger A}$ is the trace norm, and $\rho^{T_B}$ is the partial transpose with respect to a sub-region $B$, which is given as $\langle \psi_A, \psi_B| \rho |\psi_A', \psi_B' \rangle = \langle \psi_A,\psi_B'|\rho^{T_B}|\psi'_A, \psi_B\rangle$ for an arbitrary orthonormal basis $\{|\psi\rangle\}$ such that $|\psi\rangle = |\psi_A\rangle\otimes |\psi_B\rangle$. The logarithmic negativity is an entanglement monotone, which means that it is non-increasing under local quantum operations and classical communication \cite{2005lognegativity}, and it is zero for all separable states, i.e., states of the form $\rho = \sum_i p_i \rho_i^A \otimes \rho_i^B$. 

\begin{figure}[tb]
\includegraphics[width=8cm, scale=1.0]{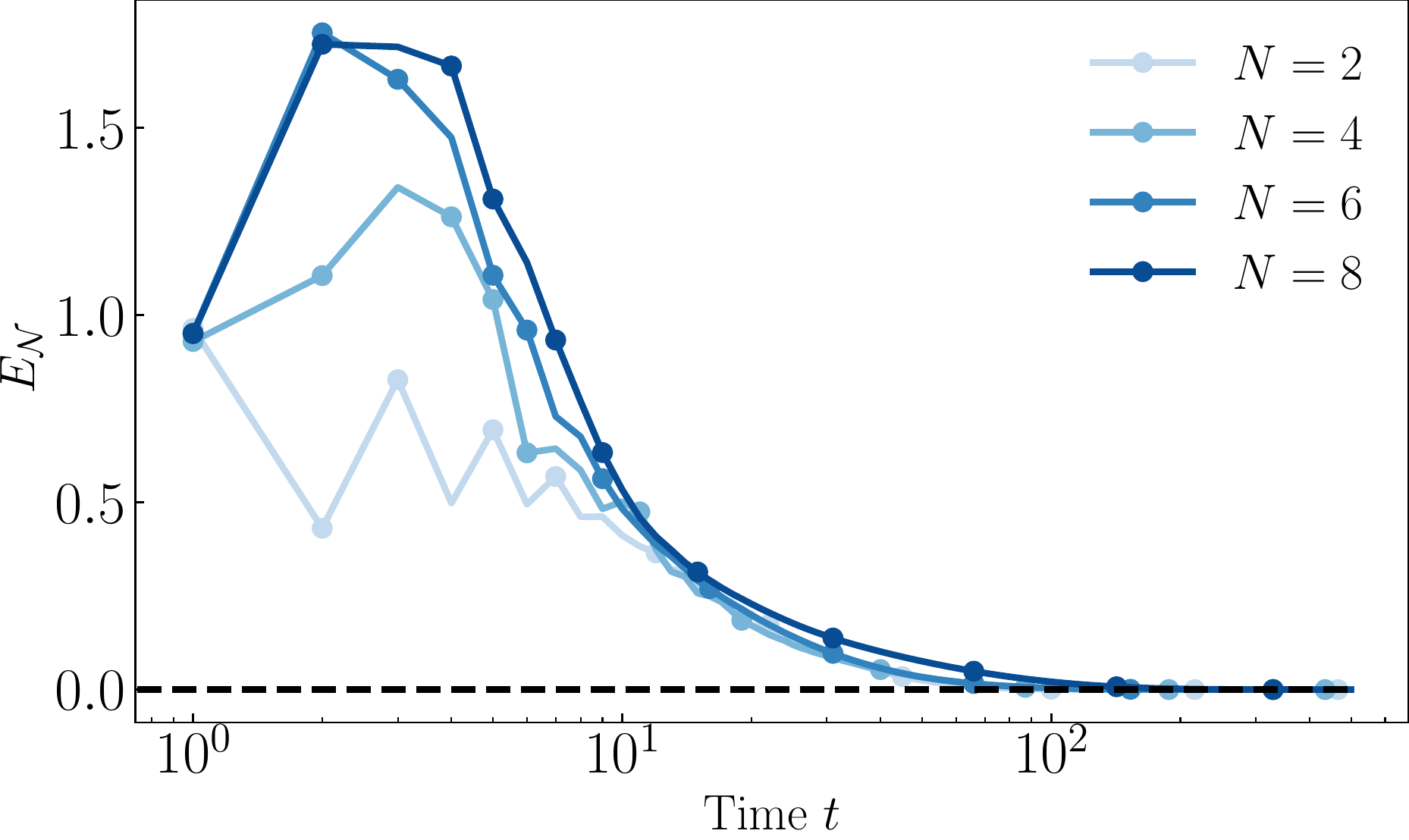}
\caption{\label{fig:Lindblad_Zj_neg} \textbf{Logarithmic negativity $E_{\mathcal{N}}$ under dephasing noise.} Time evolution of the logarithmic negativity as given in Eq.~\eqref{eq:negat} using ED, under dephasing noise $L_j = S_j^z$ with $\gamma = 0.1$. $E_{\mathcal{N}}$ increases at short times $t\lesssim 1/\gamma$, when the dephasing noise kicks in destroying quantum correlations.}
\end{figure}

We study the dynamics of $E_{\mathcal{N}}$ starting from the initial state \begin{equation} \label{eq:initial_state}
    |\psi_0\rangle = \otimes_{j=1}^N |+\rangle,
\end{equation}
which lies in the largest Krylov subspace associated to $\mathcal{C}_{\textrm{PF}}$ with dimension $\sim(2\sqrt{2})^N\approx 3^{0.95N}$~\cite{2018_PFmodel}. This corresponds to the fully-paired, i.e., trivial dot pattern, subspace.  Figure~\ref{fig:Lindblad_Zj_neg} shows the time evolution of the logarithmic negativity. At short times $t_{\text{deph}}\lesssim 1/\gamma$, $E_{\mathcal{N}}$ increases since the evolution is dominated by the unitary part. However, for $t \gtrsim t_{\text{deph}}$ the dephasing noise dominates the dynamics, destroying quantum correlations and leading to a vanishing $E_{\mathcal{N}}$. While our numerical simulations suggest that in the presence of conserved quantities $E_{\mathcal{N}}$ has a slow decay, we leave a more detailed analysis for future work. 

In fact, the stationary state under dephasing noise, Eq.~\eqref{eq:rho_ss_classical}, is a separable state for arbitrary initial states. It is the sum of projectors onto product states in the local $z$ basis $|\psi_{\alpha\beta}\rangle = |\psi_{\alpha\beta}^{A}\rangle \otimes |\psi_{\alpha\beta}^{B}\rangle$, appearing as a result of the classical fragmentation and Hermitian jump operators. Hence, it can be written as $\rho_{\textrm{ss}} = \sum_{\alpha\beta} p_{\alpha} \rho_{\alpha\beta}^{A}\otimes \rho_{\alpha\beta}^{B}$, with $p_{\alpha} = c_\alpha/D_\alpha$, and $\rho_{\alpha\beta}^{A(B)} = |\psi_{\alpha\beta}^{A(B)}\rangle\langle \psi_{\alpha\beta}^{A(B)}|$. Therefore, the logarithmic negativity for an arbitrary bipartition with an arbitrary initial state is zero. This result generalizes to stationary states for systems with Abelian commutants spanned by a local product basis and Hermitian jump operators.

\subsection{Operator space entanglement}
While quantum correlations eventually vanish in the presence of dephasing noise, information continues its spreading in the presence of conserved quantities. 
We characterize this spreading using the operator space entanglement (OSE), which measures the von Neumann entropy of the vectorized density operator $\rho \rightarrow |\psi(\rho)\rangle$, using Choi’s isomorphism $|\sigma_i\rangle\langle\sigma_i^\prime| \rightarrow |\sigma_i \sigma_i^\prime\rangle$~\cite{1975_Choi}.
With the Schmidt decomposition of $|\psi(\rho)\rangle$, the OSE is given by,
\begin{equation}
    S_{\text{OP}} = -\sum_a \lambda_a^2 \log \lambda_a^2,
\end{equation}
where the Schmidt values $\lambda_a$ are normalized to $\sum_a{\lambda_a^2} = 1$. 
In the presence of conserved quantities, the OSE can be split into two types of entanglement: the number entanglement $S_\text{num}$ and the symmetry-resolved entanglement $S_{\text{res}}$~\cite{2019_MBL_S_res_experiment, 2022_OSE_rise_and_fall},
\begin{equation}
    S_{\text{OP}} = S_{\text{num}} + S_{\text{res}}.
\end{equation}
$S_{\text{num}}$ is the Shannon entropy associated with the fluctuations of the conserved quantities in half of the system and $S_{\text{res}}$ the weighted von Neumann entanglement entropy within each symmetry sector.

\begin{figure}[tb]
\includegraphics[width=8cm]{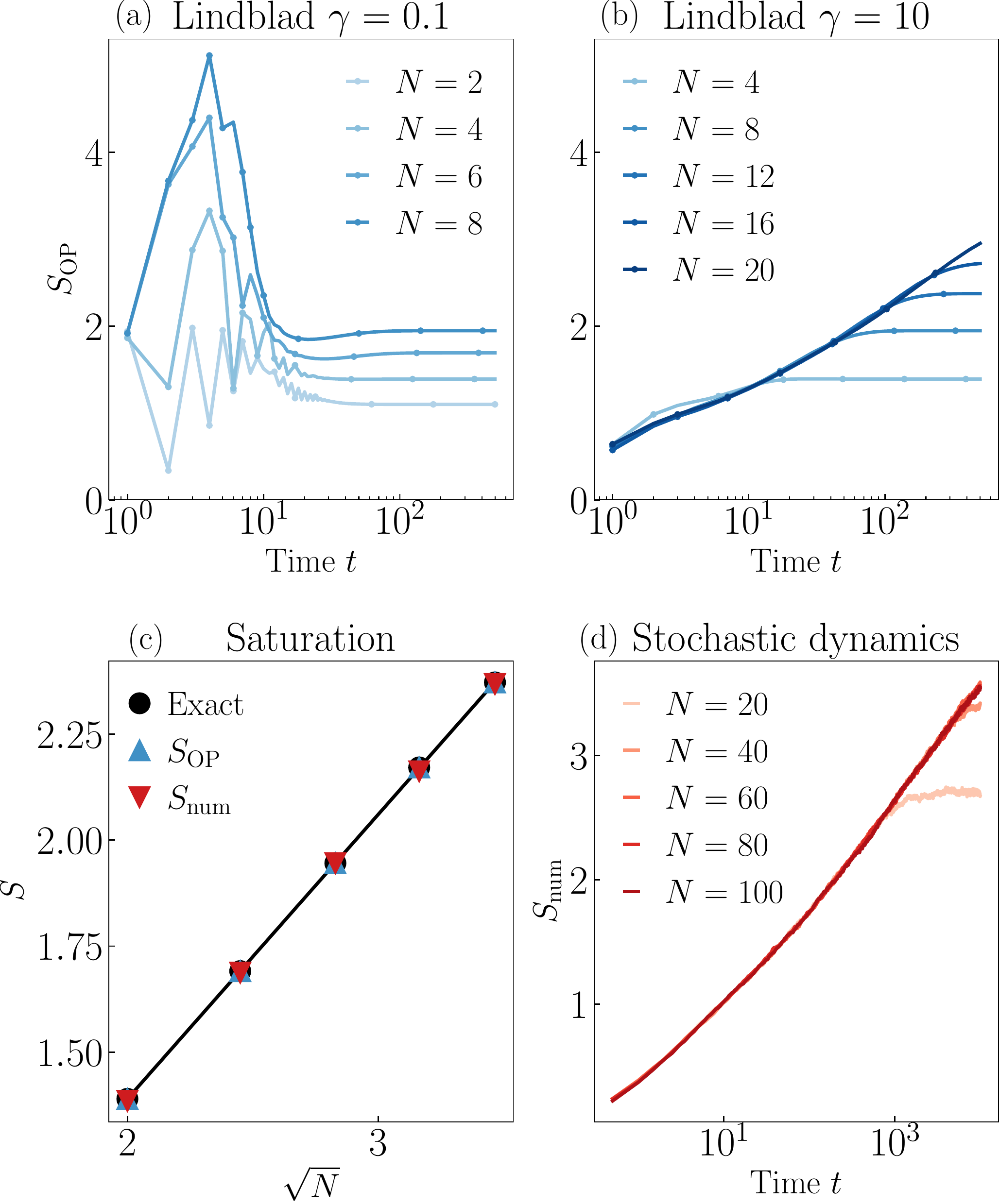}
\caption{\label{fig:Lindblad_deph_OSE_large} \textbf{Operator space entanglement and number entanglement under dephasing noise.} The initial state is $|\psi_0\rangle = \otimes_j |+\rangle_j$. (a) Lindblad dynamics of the OSE under dephasing noise $L_j = S_j^z$ with $\gamma = 0.1$ using ED. For small $\gamma$, the OSE increases at short times $t\lesssim1/\gamma$ when the dynamics is governed by the unitary term, then decreases and saturates to a size-dependent value. (b) Lindblad dynamics with large $\gamma = 10$ using TEBD. The OSE is largely suppressed by the dissipation, which allows efficient TEBD simulation. The data suggests a logarithmic growth with a rate increasing over time (see main text). (c) The analytic results of the OSE for the stationary state (black dots), the saturation values of $S_{\text{OP}}$ under Lindblad dynamics (upper-pointing triangles), and $S_{\text{num}}$ under stochastic dynamics (down-pointing triangles) show quantitative agreement. The saturation values under Lindblad dynamics are obtained with the same TEBD parameters as in (b). The OSE of the stationary state in Eq.~\eqref{eq:rho_ss_classical} scales as $O(\sqrt{N})$ with system size. (d) Number entanglement of the effective stochastic dynamics, which shows similar behavior as in the Lindblad dynamics with large $\gamma$. Each curve is averaged over $10000$ random samples.}
\end{figure}

We study the evolution of OSE starting from the same initial state $|\psi_0\rangle = \otimes_j |+\rangle_j$ in Fig.~\ref{fig:Lindblad_deph_OSE_large}. 
For small $\gamma = 0.1$, similarly to the logarithmic negativity, the OSE grows for a time $t\lesssim 1/\gamma$, and is then suppressed by the dissipation. 
However, the OSE saturates to size-dependent finite values (Fig.~\ref{fig:Lindblad_deph_OSE_large}a). 
For large $\gamma = 10$, the OSE is largely suppressed, which allows for efficient TEBD simulation for larger system sizes. 
We observe that the OSE grows even with the presence of dissipation until saturation (Fig.~\ref{fig:Lindblad_deph_OSE_large}b).
The saturation values can be calculated from the expression for the stationary state $\rho_\textrm{ss}$ in Eq.~\eqref{eq:rho_ss_classical}. 
Vectorizing the stationary density matrix $\rho_\textrm{ss}\to |\psi_\textrm{ss}\rangle$, one finds that the saturation value of the OSE is given by the von Neumann entropy of the state $|\psi_{\text{ss}}\rangle$, which was analytically obtained in Ref.~\cite{2018_PFmodel}. 
In particular, it was shown that $S_{\text{OP}}(\rho_\textrm{ss})=S_{\textrm{num}}(|\psi_{\text{ss}}\rangle)$ scales as $O(\sqrt{N})$ with system size $N$ and that $S_{\text{res}}(|\psi_{\textrm{ss}}\rangle) = 0$ (Fig.~\ref{fig:Lindblad_deph_OSE_large}c).

A recent study argued that the OSE grows logarithmically in the presence of a U$(1)$ charge validating this expectation for certain systems~\cite{2022_OSE_rise_and_fall}.
For the U$(1)$-conserving $XXZ$ chain considered in Ref.~\cite{2022_OSE_rise_and_fall}, the authors found that the strongly dephased dynamics can be approximated by a symmetric simple exclusion process of hardcore particles. There particle fluctuations across the bipartition resulted in a logarithmic growth of the number entropy, while the symmetry-resolved entanglement vanished. 

In the following, we extend this analysis to the presence of the non-local conserved quantities that characterize the fragmented structure of the stationary state, which helps to understand the OSE growth observed in Fig.~\ref{fig:Lindblad_deph_OSE_large}b.
Unlike Ref.~\cite{2022_OSE_rise_and_fall}, the number entropy of the systems we consider in this work is related to the fluctuations of non-local conserved quantities, the color-dot patterns. Analogously to the U$(1)$ charge $N_c$ that admit the decomposition $N_c=N_L+N_R$, we split the global dot pattern $A_k$ into left and right patterns such that we can keep track of their fluctuations. For example, for the fully-paired state $|\redbluetwodimer\rangle$, the left and right dot patterns after a half-chain bipartition are given by $A_k = \{\reddot\ \bluedot \}$ and $\bar{A}_k = \{\bluedot\  \reddot \}$, respectively.
Similar to the case of zero total charge with $N_R = -N_L$, the right dot pattern is a reflection of the left dot pattern for the fully-paired subspace.
As a result, the half-chain number entanglement entropy is given by
\begin{equation} \label{eq:Snum}
    S_{\text{num}} = - \sum_{A_k} p_{A_k} \log p_{A_k},
\end{equation}
with $p_{A_k}$ the probability of having the left dot pattern $A_k$~\footnote{The finest block structure in the half-chain bipartition is labeled by the left dot patterns, as it is the case in the full Hilbert space.}.

In the limit of strong dephasing, we can derive an effective Lindblad evolution using degenerate perturbation theory for open quantum systems~\cite{2012SW_for_dissipation}. We do so by splitting $\mathcal{L} = \mathcal{L}_0 + \mathcal{L}_1$ into the unperturbed contribution $\mathcal{L}_0$ and the perturbation $\mathcal{L}_1$ in the limit $|J_j|/\gamma \to 0$. Here,
\begin{equation}
    \begin{aligned}
    \mathcal{L}_0 (\rho) &= \gamma \sum_j \left( S_j^z \rho S_j^z - \frac{1}{2} \{(S_j^z)^2, \rho\}\right),\\
    \mathcal{L}_1 (\rho) &= -i[\sum_j J_j e_{j,j+1}, \rho].
    \end{aligned}
\end{equation}
Since the initial state $|\psi_0\rangle = \otimes_j |+\rangle_j$ lies in the fully-paired subspace of the PF model, the stationary states of $\mathcal{L}_0$ are given by  $\rho_0^{\boldsymbol{\sigma}} = |\boldsymbol{\sigma}\rangle\langle\boldsymbol{\sigma}|$, where $|\boldsymbol{\sigma}\rangle$ are all possible fully-paired product states. The perturbation $\mathcal{L}_1$ breaks this degeneracy inducing transitions among different $\rho_0^{\boldsymbol{\sigma}}$. 
Performing the perturbation theory to second order in $|J_j|/\gamma$ we find the effective Liouvillian~\cite{2012SW_for_dissipation, 2013_effective_dephasing, 2016_Dephasing_MBL, 2022_OSE_rise_and_fall}
\begin{equation}\label{eq:eff_Lindblad_second_order_perturbation}
    \mathcal{L}_{\text{eff}} = -\mathcal{P} \mathcal{L}_1 (\mathcal{L}_0)^{-1}\mathcal{L}_1 \mathcal{P},
\end{equation}
where $\mathcal{P}$ is the projection onto the subspace spanned by $\rho_0^{\boldsymbol{\sigma}}$. 
This effective dynamics reduces to a classical Markov evolution $\partial_t \rho(t) = -\mathbb{W}_{\textrm{eff}}\rho(t)$ for the diagonal components of $\rho$ in the fully-paired product basis $\rho_0^{\sigma}$ with 
\begin{equation}
    \mathcal{L}_{\text{eff}} (\rho_0^{\boldsymbol{\sigma}}) = -\sum_{\boldsymbol{\sigma}'}  \langle \boldsymbol{\sigma}' | \mathbb{W}_{\text{eff}}|\boldsymbol{\sigma}\rangle  \rho_0^{\boldsymbol{\sigma}'}. 
\end{equation}
$\mathbb{W}_{\text{eff}} = \sum_j g_j^{\alpha\beta} (|\alpha \alpha\rangle\langle \beta\beta|)_{j,j+1}$ is the Markov generator given by a PF model with coefficients $g_j^{\alpha\beta}$ obtained in App.~\ref{subapp:SW_Lindblad}. This implies that the effective dynamics indeed preserves the commutant algebra associated to the PF model $\mathcal{C}_{\textrm{PF}}$.

For an XXZ model under dephasing noise in Ref.~\cite{2022_OSE_rise_and_fall}, the corresponding effective stochastic evolution can be mapped to a simple exclusion process, from where an analytical prediction for the growth of $S_{\textrm{num}}$ could be obtained. However, we are not aware of any analysis of the evolution generated by $\mathbb{W}_{\text{eff}}$. Hence, we numerically simulate it in a manner that can be compared to the implementation for open quantum dynamics. In the basis spanned by $\{\rho_0^\sigma\}$, the probability vector with entries $p_\sigma(t)$ at discrete time $t$ is given by $p_\sigma(t)=\sum_{\sigma^\prime} (P^t)_{\sigma \sigma^\prime} p_{\sigma^\prime}(0)$ where $P = e^{-\mathbb{W}_{\text{eff}}}$~\cite{2016_Dephasing_MBL}. Transition probabilities are given by the corresponding entry in the matrix $P$, which is symmetric, and satisfies $P_{\sigma \sigma^\prime}\in [0, 1]$ together with $\sum_\sigma P_{\sigma \sigma^\prime} = 1$. Hence, detailed balance holds with respect to a stationary state, which is the uniform distribution over all fully-paired states. This corresponds to the stationary state $\rho_{\text{ss}}$ of the Lindblad dynamics. To efficiently implement this evolution, we consider a brick-wall circuit structure where $2$-site local gates $P_{j,j+1}$ randomly permute among two-site local spin configurations in the $z$-basis as, e.g., in Refs.~\cite{2020_Pablo_automato, Morningstar_2020, Iaconis_2019, Iaconis_2021,Hart_2022, feldmeier2021critically,2022_Lehmann_Pablo_Markov}. Starting from the initial product state $\otimes_j|+\rangle_j$, we then compute the evolution of the number entropy $S_{\text{num}}$ as given in Eq.~\eqref{eq:Snum} by averaging over various circuit realizations. More details about the numerical implementation can be found in App.~\ref{subapp:stochastic_dynamics}.  

In Fig.~\ref{fig:Lindblad_deph_OSE_large}, we compare the open quantum dynamics (panel b with $\gamma=10$) with the stochastic one in panel d. 
The latter allows us to simulate larger system sizes and longer times than what is accessible by TEBD simulations. 
We observe a growth of the number entanglement of the stochastic model in Fig.~\ref{fig:Lindblad_deph_OSE_large}b, which agrees with the numerical results obtained in the quantum setup. 
However, we are unable to provide an analytical prediction for the observed scaling of growth as for the U$(1)$-symmetric systems.
Assuming a logarithmic growth of the OSE $S(t) = S_0 + \eta\log(t)$, we find that the growth rate $\eta$ slightly increases over time.
Note that a similar effect is also observed in Fig.~2a of Ref.~\cite{2022_OSE_rise_and_fall} for U$(1)$ symmetric systems, which is caused by finite time effects. 
Our numerical simulations reach a saturation value for the $S_\mathrm{num}$ (red down-pointing triangles) that agrees with the analytical result (black dots) and the saturation of the OSE under the quantum Lindblad dynamics (blue upper-pointing triangles) as shown in Fig.~\ref{fig:Lindblad_deph_OSE_large}c.

\section{Structure-preserving noise}\label{sec:preserving}
In the previous section, we observed that the dephasing noise reduced the QF of the TL model to the classical one. This led to vanishing quantum correlations as measured by the $E_{\mathcal{N}}$, while classical correlations ($S_{\textrm{num}}$) could still propagate due to fluctuations of the remaining conserved quantities. We now consider a dissipative bath preserving the QF and investigate the effects of the system being fragmented in an entangled basis. We choose $L_j = e_{j,j+1}$ acting on two consecutive sites, which is an element of the bond algebra $\mathcal{A}_{\textrm{TL}}$. Hence, the open commutant algebra agrees with that of the TL model $\mathcal{C}^O = \langle\{h_i\}, \{L_j\}\rangle' = \mathcal{C}_{\text{TL}}$. 

\subsection{Stationary states with quantum fragmentation}
\begin{figure}[bt]
\includegraphics[width=8cm]{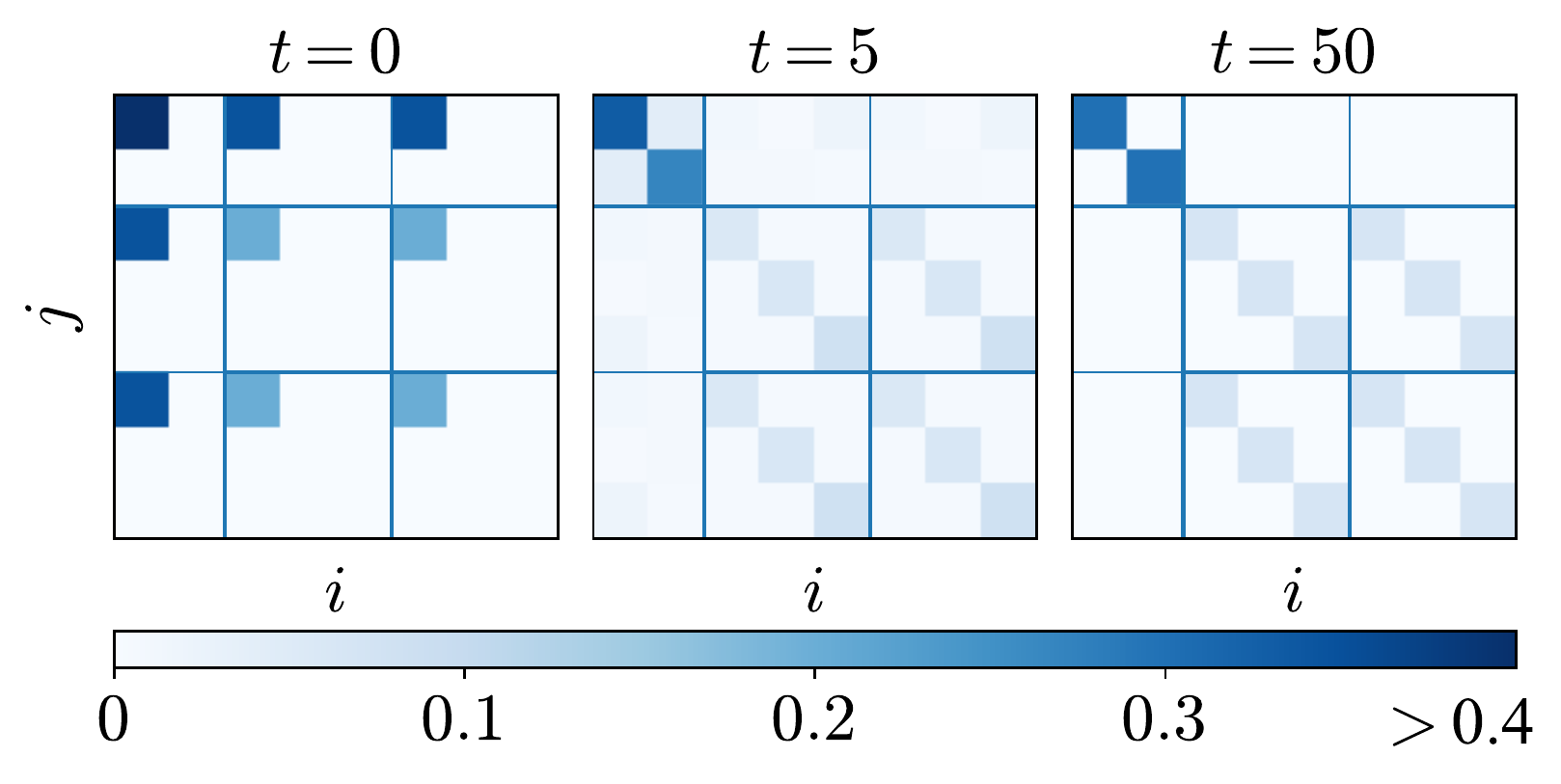}
\caption{\label{fig:TL_rho_ej} \textbf{Time evolution of the density matrix under structure-preserving noise.} Time evolution of the density operator under structure-preserving noise $L_j = e_{j,j+1}$ using ED, with system size $N=4$ and $\gamma=1$. The color indicates the matrix elements $|\rho_{ij}|$ in the entangled basis. The stationary state consists of projected identities in all diagonal blocks and degenerate off-diagonal blocks, while all elements in non-degenerate blocks vanish.}
\end{figure}

When considering quantum structure-preserving noise, the stationary state inherits the QF of the TL model leading to the general expression
\begin{equation}\label{eq:rho_ss_quantum}
	\rho_{\text{ss}}  = \sum_{\lambda,\alpha,\alpha^\prime}  (M_{\lambda})_{\alpha \alpha^\prime}\frac{\Pi_{\alpha\alpha^\prime}^\lambda\mathbb{1}}{D_\lambda}= \bigoplus_{\lambda} \left( M_{\lambda} \otimes \frac{\mathbb{1}_{\lambda}}{D_\lambda} \right),
\end{equation}
where $(M_{\lambda})_{\alpha \alpha^\prime} = \mathrm{Tr}(\Pi_{\alpha^\prime \alpha}^\lambda\rho_0)$ is the $d_\lambda \times d_\lambda$ matrix of overlaps between the initial state $\rho_0$ and $\Pi_{\alpha{\alpha^\prime}}^{\lambda}$ with $\Pi_{\alpha}^{\lambda}\equiv\Pi_{\alpha\alpha}^{\lambda}$. 
There are two major differences which distinguishes this from the stationary state discussed in the previous section. 
First, there are stationary phase coherences, i.e., $\mathcal{L}(\Pi_{\alpha \alpha^\prime}^\lambda \mathbb{1})= 0$, captured by the non-zero overlaps with the conserved intertwine operators $\Pi_{\alpha\alpha^\prime}^{\lambda}$. Recall that these appear as a consequence of $\mathcal{C}_{\text{TL}}$ being non-Abelian. As in the case of dephasing noise, the conserved projectors give the stationary state $\Pi_{\alpha}^{\lambda}\mathbb{1}/D_\lambda$ in the diagonal subspaces. 
These projected identities indicate full decoherence within the subspaces $\mathcal{H}^{(\mathcal{A})}_\lambda$ induced by $L_j \in \mathcal{A}$.
Nonetheless, intertwine operators acting on the off-diagonal subspaces, guarantee non-vanishing coherences for generic initial states~\cite{Baumgartner_2008_2}, indicating that the whole system does not fully decohere. 
Figure~\ref{fig:TL_rho_ej} shows an example of the Lindblad evolution for the initial state in Eq.~\eqref{eq:initial_state_two_degenerate} displaying non-zero overlap onto the non-degenerate fully-dimerized subspace ($\lambda = 0$) and onto two degenerate Krylov subspaces ($\lambda = 1$) in the entangled basis of the TL model. 
The system evolves to the stationary state with projected identities both in the diagonal and off-diagonal degenerate subspaces.
Second, the projected identity $\Pi_{\alpha \alpha^\prime}\mathbb{1}$ within each Krylov subspace is a mixture of entangled basis states. This implies that the stationary state is typically not separable unless for fine-tuned initial states. As we find in the following, this is also signalled by the behavior of the logarithmic negativity. 

Moreover, the exponentially large (in system size) dimension of the commutant algebra as a consequence of HSF, $\dim (\mathcal{C}) = \sum_\lambda d_\lambda^2 \sim e^{aN}$, turns into a strong ---coherent in the case of non-Abelian $\mathcal{C}$--- memory of the initial configuration. 
Information about the initial state is stored by the weight on the invariant subspaces $\mathcal{H}^{(\mathcal{C})}_\lambda$.
When $L_j\in\mathcal{A}$, $\mathcal{H}^{(\mathcal{C})}_\lambda$ are decoherence-free subspaces and noiseless subsystems immune to dissipation, which are extensively studied in the context of error correction and fault tolerant quantum computation~\cite{1997_Zanardi_noiseless_quantum_codes, 1998_DFS_Lidar_Chuang_Whaley, 2000_Knill_noiseless_subsystem_error_correction, 2001_DFS_fault_tolerant_quantum_computation, 2003_DFS_review, 2006_Choi_NS_commutant}.

\subsection{Infinite-temperature autocorrelation function}
Once again we can use a similar analysis to that of the stationary state to derive the saturation value of the spin-spin autocorrelation function $\langle S^z_j(t)S^z_j(0)\rangle$. One finds that a general operator $O$ relaxes to the stationary value
\begin{equation}
    O(\infty) = \bigoplus_\lambda \left( O_\lambda \otimes \frac{ \mathbb{1}_\lambda }{D_\lambda}\right), 
\end{equation}
where $(O_\lambda)_{\alpha \alpha^\prime} = \mathrm{Tr}(\Pi^\lambda_{\alpha^\prime\alpha}O)$. Here, $O_\lambda$ is a $d_\lambda \times d_\lambda$ matrix with elements given by the overlap of the operator and the corresponding projector or intertwine operator.

Therefore, for a local operator $S_j^z$, the saturation value of the autocorrelation is given by $\mathrm{Tr}(S_j^z(\infty) S_j^z(0))/3^N$, which is exactly the Mazur bound of the TL model for unitary evolution
\begin{equation}\label{eq:Mazur_quantum}
    M_{\text{TL}}(S_j^z) = \frac{1}{3^N} \sum_{\lambda, \alpha, \alpha^\prime}\frac{|\mathrm{Tr}(\Pi_{\alpha\alpha^\prime}^{\lambda}S_j^z)|^2}{D_\lambda}.
\end{equation}
This agrees with the numerical results shown in Fig.~\ref{fig:Lindblad_autocorrelation}, where the autocorrelation functions saturate to the same value for the closed system (green solid line) and under the structure-preserving noise (upper-pointing triangles). The finite-size scaling of the saturation values suggests that it is not vanishing either in the bulk or at the edge (see App.~\ref{subapp:fss_ac}). 

\subsection{Logarithmic negativity and operator space entanglement}
A vanishing or non-vanishing bulk autocorrelation function is not sufficient to distinguish classical from quantum fragmentation.
For example, the bulk autocorrelation functions decay to zero for the $t-J_z$ chain but remain finite for certain dipole-conserving models, both of which are classical fragmented~\cite{2020_sala_ergodicity-breaking, 2020_SLIOMs}.

\begin{figure}[bt]
\includegraphics[width=8cm, scale=1]{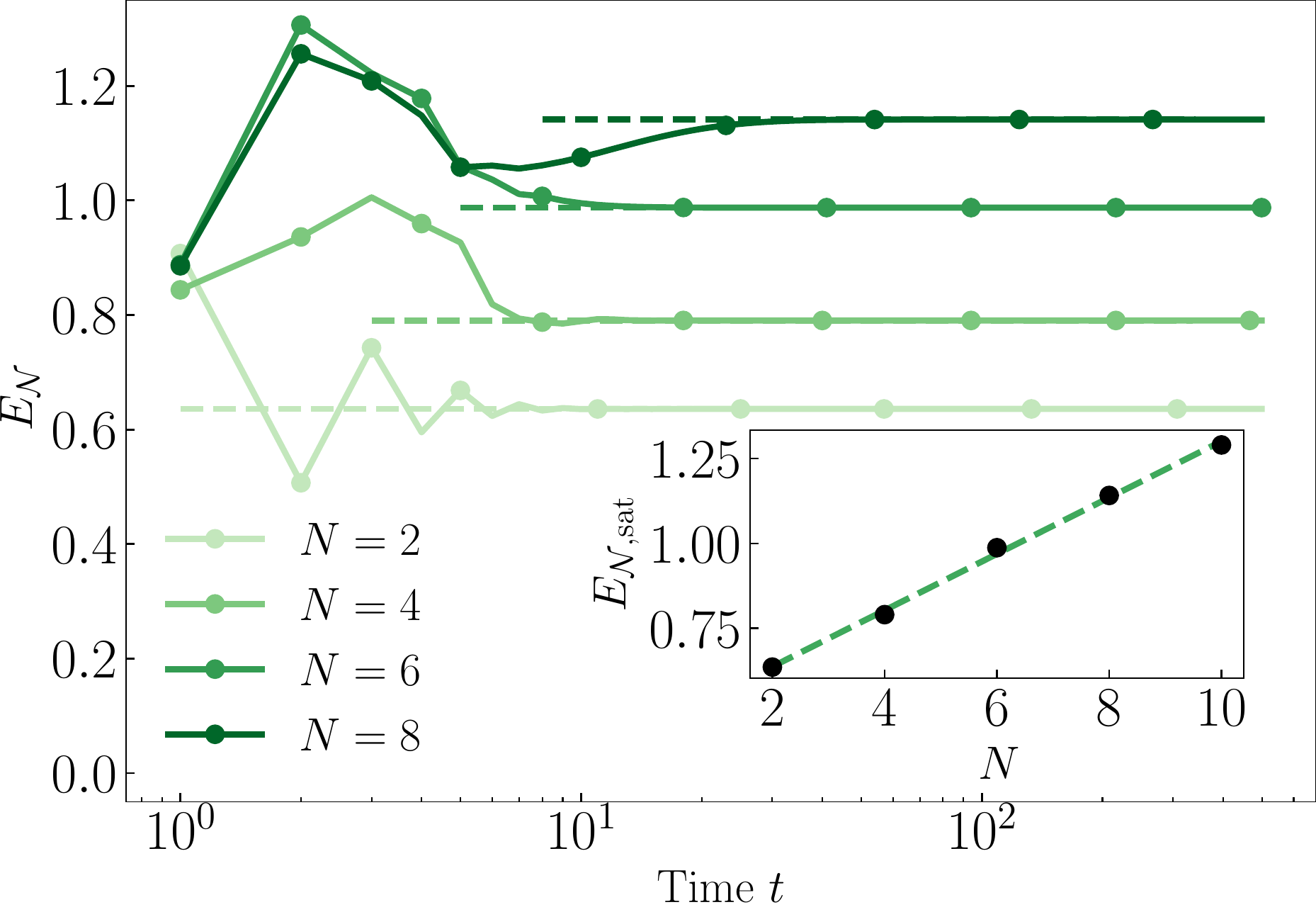}
\caption{\label{fig:Lindblad_ej_neg} \textbf{Logarithmic negativity under structure-preserving noise.} Time evolution of the logarithmic negativity using ED under the structure-preserving noise $L_j = e_{j,j+1}$, with $\gamma = 0.1$. The initial state is $|\psi_0\rangle = \otimes_j |+\rangle_j$. The logarithmic negativity $E_{\mathcal{N}}$ increases at $t\lesssim 1/\gamma$ and then saturates to a finite size-dependent value. In the inset, we show the scaling of this saturation value as directly computed from the stationary state in Eq.~\eqref{eq:rho_ss_quantum}, also included in the main panel (dashed green lines). While our numerical simulations are limited to system sizes $N\leq 10$, the scaling with system size suggests volume-law.}
\end{figure}

However, a sharp contrast can be detected in the behavior of the logarithmic negativity in the presence of different types of baths. 
While we found a vanishing negativity for dephasing noise when starting from the initial state $\otimes_j |+\rangle_j$, we find that $E_{\mathcal{N}}$ saturates to a size-dependent value at long times under the structure-preserving noise, indicating that the system evolves towards an entangled stationary state (see Fig.~\ref{fig:Lindblad_ej_neg}). 
Moreover, the scaling of the negativity with system size is directly computed from the stationary state in Eq.~\eqref{eq:rho_ss_quantum} and shown in the inset of Fig.~\ref{fig:Lindblad_ej_neg}, suggesting that the stationary state satisfies a volume-law. 
The source of this non-vanishing value is the fact that the system is fragmented in an entangled basis, hence providing a clear signature to distinguish quantum and classical fragmentation. 

Thus we propose the logarithmic negativity of stationary states as a probe to distinguish quantum from classical fragmentation. 
Generally, identifying CF structure is an easier task that can be achieved by iteratively applying local terms of Hamiltonian to a root product state. 
However, there can still be a finer structure within these Krylov subspaces due to quantum fragmentation appearing in an unknown entangled basis. 
To detect whether such a finer structure exists, one could start from an initial state within a Krylov subspace, and study the dynamics of the logarithmic negativity under a dissipative bath, which should preserve all the symmetries of the Hamiltonian. 
This means that the jump operators should be elements of the bond algebra $L_j\in\mathcal{A}$ and Hermitian. 
While systems showcasing only CF evolve towards a separable stationary state with zero negativity, systems that are quantum fragmented can lead to non-zero logarithmic negativity. 

\begin{figure}[bt]
\includegraphics[width=8cm, scale=1]{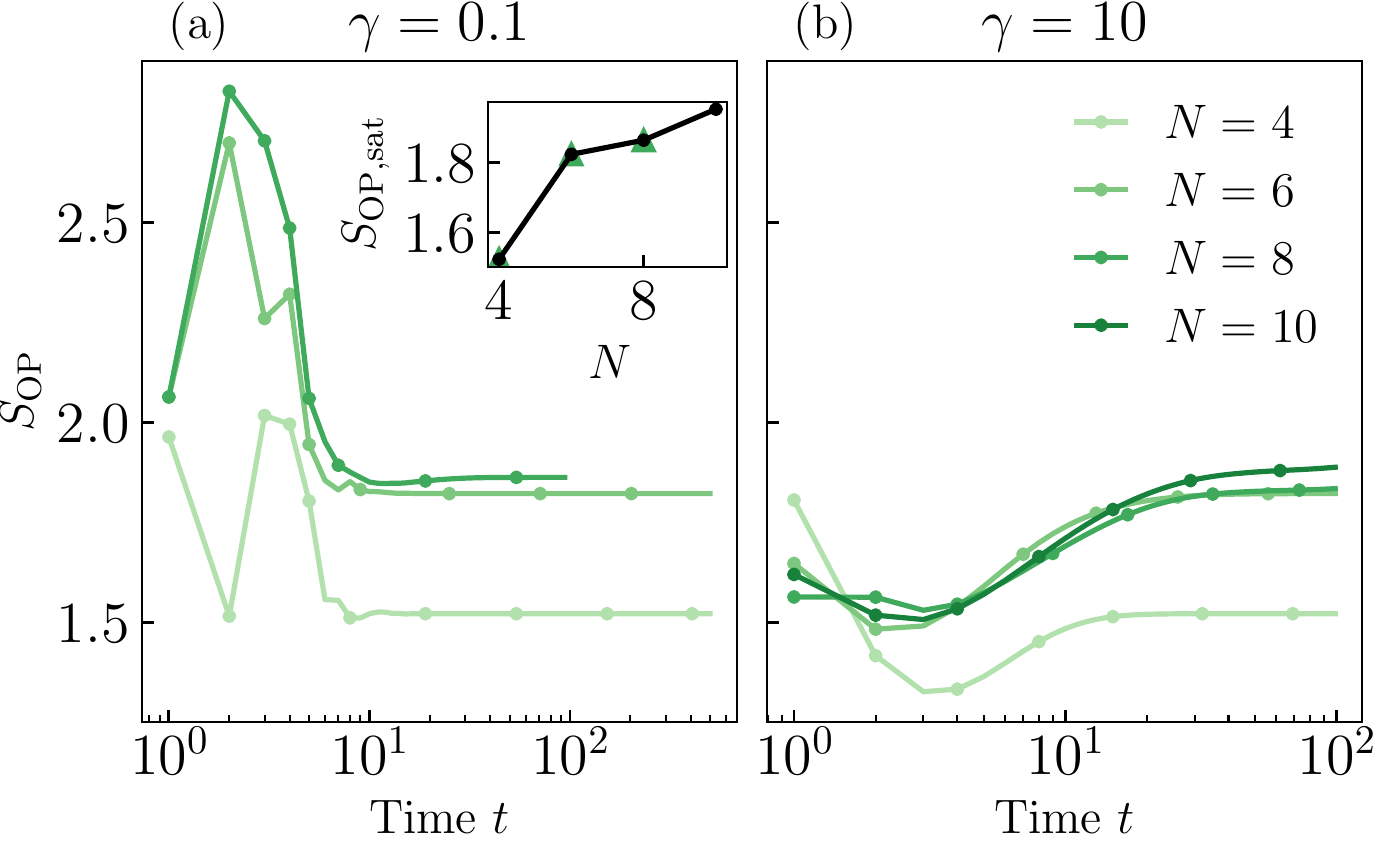}
\caption{\label{fig:Lindblad_ej} \textbf{OSE under structure-preserving noise.} Time evolution of the OSE under the structure-preserving noise $L_j = e_{j,j+1}$ with the initial state $|\psi_0\rangle = \otimes_j |+\rangle_j$. Data is obtained using ED for $\gamma = 0.1$ and TEBD for $\gamma = 10$ with maximal bond dimension $\chi=1000$. (a) For $\gamma = 0.1$, the OSE increases at short times and then decreases and saturates to a finite value. The inset shows the OSE of the stationary state obtained from Eq.~\eqref{eq:rho_ss_quantum} (circle), which matches the saturation values obtained by ED (upper-pointing triangle). (b) For $\gamma = 10$, the OSE saturates to the same values as for $\gamma = 0.1$.}
\end{figure}

Before concluding this section, we study the evolution of the OSE and compare its saturation value to that obtained from the stationary state. 
The results are shown in Fig.~\ref{fig:Lindblad_ej} for $\gamma=0.1$ (panel a) and $\gamma=10$ (panel b). 
In this case, the dynamics of OSE cannot be efficiently studied even in the regime $\gamma \gg 1$ for the following reasons. 
First, the stationary-state subspace of the unperturbed contribution $\mathcal{L}_0$ is spanned by entangled states. 
Obtaining an orthonormal set of these entangled states requires full diagonalization of $\mathcal{L}_0$. 
Second, transition among entangled states cannot be modelled by local updates on local configurations and hence cannot be mapped to a classical stochastic circuit evolution. 
Moreover, while under dephasing noise we could directly extract the number entanglement by calculating the probabilities of the dot patterns of the product states, this is not the case for entangled states which involved entangled dot patterns. 
This raises the general question whether one can capture quantum fragmentation phenomena using classical stochastic dynamics.

\section{Conclusions and outlook}\label{sec:conclusion}
The goal of our work was to examine how HSF impacts open Lindblad dynamics, taking into account whether the coupling to the bath maintains or disrupts fragmentation in an entangled basis. By analyzing the symmetries of the Liouvillian, we were able to analytically derive the stationary state and characterize the dynamics of autocorrelation functions and entanglement combining analytical and numerical methods. First, we found that for a dephasing noise ---that reduces the quantum fragmentation of the TL model to the classical fragmentation of the PF--- the stationary state is a separable state with zero quantum correlations. This holds generically for classically fragmented open systems with Hermitian jump operators. Nonetheless, the OSE increases as a function of time due to the fluctuations of the non-local conserved charges as captured by an effective stochastic evolution in the regime of strong dephasing. On the other hand, for a dissipative coupling preserving the QF of the TL model, the system evolves to a highly-entangled stationary state with size-dependent logarithmic negativity. This finite saturation value is a dynamical property distinguishing classical from quantum fragmentation in open quantum systems, while for unitary evolution both classical and quantum fragmentation lead to volume-law entanglement entropies. In addition, there exist stationary coherences in the off-diagonal subspaces due to non-Abelian commutant algebras, indicating that the system does not fully decohere. Although the system shows distinct entanglement properties under the two couplings, finite autocorrelation functions could persist under both types of dissipation. Moreover, the extensive fragmentation of the Hilbert space translates into exponentially many (in the volume of the system) degenerate stationary states signaling a strong dependence on the initial state. 

The preceding discussion has highlighted three critical components: 
(1) the distinction between classical and quantum fragmentation, which is synonymous with product and entangled basis spanning the fragmented structure respectively.
This translates into stationary identity matrices within Krylov subspaces in terms of either product or entangled states, respectively, where the latter leads to a finite negativity at long times.
(2) The distinction between Abelian and non-Abelian commutants; A non-Abelian commutant results in the presence of stationary coherences, which indicates a coherent memory of the initial state~\cite{2003_Kuperberg_hybrid_quantum_memeory,2014_Albert};
and (3) the exponential dimension of the commutant as caused by HSF, which leads to a large degeneracy of stationary states and a strong dependence on the initial state.

For future work, it will be interesting to understand whether similar entanglement dynamics as the one found for quantum fragmented systems, appears for polynomially large commutants. For example, conventional symmetries such as SU$(2)$ also lead to a decomposition of the Hilbert space into symmetry sectors spanned by an entangled basis, which may evolve to a stationary state with finite negativity for specific initial states. However, with exponentially large subspaces that scale as the size of the Hilbert space, the stationary state is highly mixed, which can exhibit a different dependence of entanglement with system size.

We also leave it open to explore classical and quantum fragmentation in the presence of weak symmetries~\cite{2012_Buca_Prosen, 2014_Albert}. In fact, an example of classical (local) fragmentation in this weak sense already appeared in Ref.~\cite{2020_essler_open_fragmentation}. A natural adaptation of the commutant algebra formalism consists of considering the vectorized form of the Lindbladian $\mathcal{L}\to \tilde{\mathcal{L}}$ acting on the Hilbert space $\mathcal{H}\otimes \mathcal{H}$ and define the commutant as the set of (super)-operators commuting with every local term of $\tilde{\mathcal{L}}$. For example, it would be interesting to understand whether there are examples of quantum fragmentation and non-Abelian commutants for weak symmetries, and if so, whether they lead to similar phenomenology as the one found in this work. 

Finally, while several recent studies~\cite{Iaconis_2019,2020_Pablo_automato,  2022_Frey_HSF_fermihubbard_Markov, Morningstar_2020, Iaconis_2021,Hart_2022, feldmeier2021critically,2022_Lehmann_Pablo_Markov,2023_Feng_HSF_boson_weak_strong_transition} have employed block (local) cellular automaton dynamics to investigate the impact of classical fragmentation on infinite-temperature correlations, our work raises the following question: 
Is it possible to construct a blocked cellular automaton with finite-size gates that simulates the dynamics and capture the entanglement properties of quantum fragmentation? 
If it is not possible, the obstruction to find such cellular automaton could be used as a definition of quantum fragmentation.

\begin{acknowledgments}
The tensor-network calculations in this work were performed using the TeNPy Library~\cite{2018_Tenpy}. The authors thank Barbara Kraus, Olexei Motrunich and Sanjay Moudgalya for valuable advice. Y. L. thanks Fabian Essler, Zongping Gong, Johannes Hauschild, Dieter Jaksch, Yujie Liu, Hannes Pichler, Elisabeth Wybo, and Zhongda Zeng for helpful discussions. P.S. acknowledges support by the Walter Burke Institute for Theoretical Physics at Caltech, and the Institute for Quantum Information and Matter. This research was financially supported by the European Research Council (ERC) under the European Union’s Horizon 2020 research and innovation program under grant agreement No.~771537. F.P. acknowledges the support of the Deutsche Forschungsgemeinschaft (DFG, German Research Foundation) under Germany’s Excellence Strategy EXC-2111-390814868. F.P.’s research is part of the Munich Quantum Valley, which is supported by the Bavarian state government with funds from the Hightech Agenda Bayern Plus.
\end{acknowledgments}

\textbf{Data and materials availability.}
Data analysis and simulation codes are available on Zenodo upon reasonable request~\cite{Data_set}.

\appendix
\section{Fragmentation of PF model and TL model}
\subsection{Entangled fragmentation basis of the TL model}\label{subapp:TL_basis}
The TL model exhibits QF (in an entangled basis) where the Krylov subspaces are labeled by product or entangled dot patterns. In addition, due to the non-Abelian commutant algebra $\mathcal{C}_{\text{TL}}$, the Krylov subspaces with dot patterns of the same length are degenerate. 

We provide some simple examples of how to construct the entangled basis of the TL model. We label the basis states by $|\psi^{\lambda}_{\alpha\beta}\rangle$, where $2\lambda$ is the number of dots, $\alpha = 1, ..., d_\lambda$ denotes different degenerate Krylov subspaces for fixed $\lambda$, and $\beta$ denotes different basis states in the same Krylov subspace. For a system with two sites $N=2$, the fully-dimerized Krylov subspace with $\lambda =0$ (zero dots) is one dimensional, with $|\psi^{0}_{1, 1}\rangle = |\dimer\rangle$. For $\lambda = 1$ with two dots, the Krylov subspaces are also one-dimensional with $|\psi^{1}_{\alpha, 1}\rangle = |\onedot\,\,\,\onedot\rangle$, such that $e_{j,j+1}|\onedot\,\,\,\onedot\rangle = 0$. The dot state can be a product state, $|\sigma_1 \sigma_2\rangle$ with $\sigma_1 \neq\sigma_2$, or an entangled state such as $\frac{1}{\sqrt{2}}(|++\rangle - |--\rangle)$. The Krylov subspaces with $N=2$ and $\lambda=1$ have a degeneracy of $d_1=8$, i.e., there are in total eight different dot patterns which consist of two dots. Note that the choice of dot patterns is not unique, any linear superposition of dot patterns works. For larger system sizes with Krylov subspaces of dimension $D_\lambda \geq 1$, we apply $e_{j,j+1}$ on a root state of the subspace to generate other basis states. For example, for $N=4$ with $\lambda =1$, the Krylov subspace is three-dimensional with basis states $|\psi_{\alpha 1}^1\rangle = |\dimer\,\,\,\onedot\,\,\,\onedot\rangle$, $|\psi_{\alpha 2}^1\rangle = e_{1, 2}|\psi_{\alpha 1}^1\rangle = |\onedot \,\,\,\dimer \,\,\,\onedot\rangle$ and $|\psi_{\alpha 3}^1\rangle = e_{2, 3}|\psi_{\alpha 2}^1\rangle = |\onedot \,\,\,\onedot \ \ \ \dimer\rangle$. The dot pattern $(\onedot\,\,\,\onedot)$ is conserved and labels this Krylov subspace. A systematic way to construct the complete basis is given by Ref.~\cite{2010_TL}. 

\subsection{Finite-size scaling of the autocorrelation functions}\label{subapp:fss_ac}
\begin{figure}[bt]
\includegraphics[width=8cm]{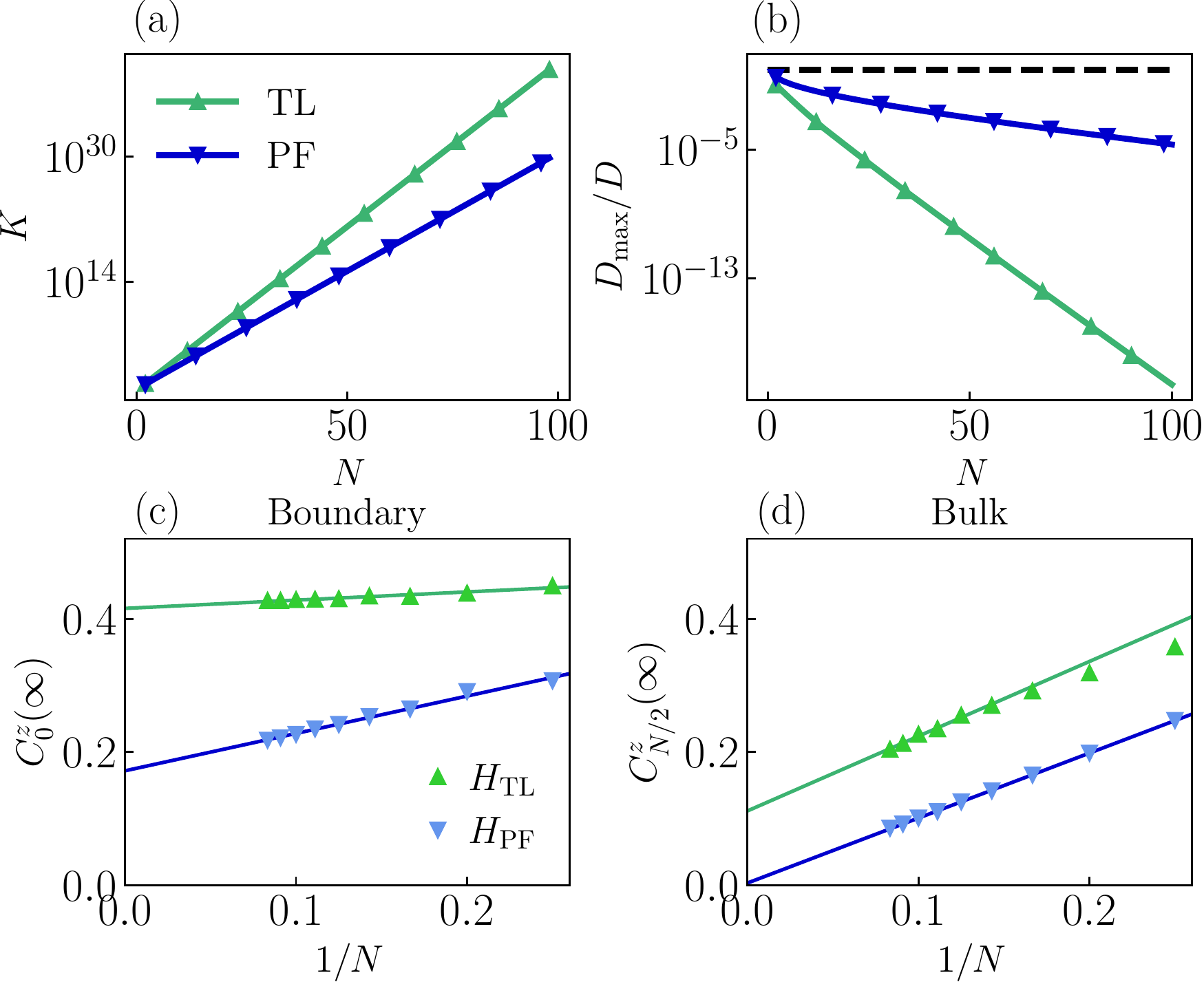}
\caption{\label{fig:TL_ED_autocorre_sat} \textbf{Strong fragmentation and saturation of autocorrelation functions.} (a) Number of Krylov subspaces $K = \sum_\lambda d_\lambda$ for PF and TL model, which scales exponentially with the system size. (b) Ratio between the dimension of the largest Krylov subspace $D_\text{max}$ and the dimension of the total Hilbert space $D=3^N$, which scales as $D_{\text{max}}/D\sim \exp (-aN)$ with $a<1$. This indicates that both models exhibit strong fragmentation. (c-d) Finite-size scaling of the long-time average of the autocorrelation functions at the boundary ($j = 0$) and at the bulk ($j = N/2$) as $1/N$. At the boundary, both the PF and TL model show infinite coherence time in the thermodynamic limit. At the bulk, the TL model has non-vanishing autocorrelation functions, while for the PF model, the autocorrelation function decays as $1/N$ and vanishes.}
\end{figure}

Both the PF and TL models exhibit strong fragmentation~\cite{2020_sala_ergodicity-breaking, moudgalya_hilbert_2022}. Figure~\ref{fig:TL_ED_autocorre_sat}a shows that the number of Krylov subspaces for the PF and the TL models scales exponentially with the system size~\cite{2018_PFmodel, 2010_TL}. Figure~\ref{fig:TL_ED_autocorre_sat}b shows that the ratio between the dimension of their largest Krylov subspace and the total Hilbert space dimension scales as $D_{\text{max}}/D\sim \exp (-aN)$.

We study the non-ergodic behavior due to strong fragmentation with the long-time average of autocorrelation functions, which is given by
\begin{equation}
    C^z_j(\infty) =  \lim_{T\rightarrow\infty} \frac{1}{T} \int_0^T dt \langle S_j^z(t) S_j^z(0)\rangle_c.
\end{equation}
We study $C^z_{j}(\infty)$ with random circuits using ED, which is shown in Fig.~\ref{fig:TL_ED_autocorre_sat}(c-d). At the boundary, $C^z_0(\infty)$ decays with the system size for both TL and PF model, but saturates to a finite value in the thermodynamic limit. This indicates that there are infinite coherence times at the boundary for both classical and quantum fragmentation. At the bulk, as discussed in the main text and in Fig.~\ref{fig:Lindblad_autocorrelation}, the autocorrelation functions $C^z_{N/2}(\infty)$ of the PF and TL models coincide with the saturation values in open systems under dephasing noise and the structure-preserving noise, respectively. The bulk autocorrelation decays as $1/N$ and vanishes for the PF model, while for the TL model, the numerical results suggest that the autocorrelation functions saturate to finite values.

\section{Long-time behaviors under Lindblad dynamics}
\subsection{The open commutant algebra under dephasing noise}\label{subapp:vonNeumann_deph} 
Here we prove that the dynamics of the TL model under dephasing noise characterized by $\mathcal{C}_{\mathrm{deph}}^O = \langle \{e_{i,i+1}\}, \{S_j^z\}\rangle^\prime$ is exactly $\mathcal{C}_{\mathrm{PF}}$. First, $e_{i,i+1}$ and $S_j^z$ are elements of $\mathcal{A}_{\mathrm{PF}}$, therefore, $\mathcal{A}_{\mathrm{deph}}^O \subseteq \mathcal{A}_{\mathrm{PF}}$. Second, all elements of the PF algebra can be generated by the local terms of the TL model and the dephasing noise. For example, the local term $(|++\rangle\langle++|)_{j,j+1} = \frac{1}{4} [S_j^z + (S_j^z)^2] e_{j,j+1} [S_j^z + (S_j^z)^2]$. The linear combinations of such products of $S_j^z$ and $e_{i,i+1}$ produce all the local terms of the PF model, which indicates that $\mathcal{A}_\mathrm{PF} \subseteq \mathcal{A}_{\mathrm{deph}}^O$. Hence, we have $\mathcal{A}_{\mathrm{deph}}^O = \mathcal{A}_{\mathrm{PF}}$, which also means that they have the same commutant, $\mathcal{C}^O_{\mathrm{deph}} = \mathcal{C}_{\mathrm{PF}}$. Altogether one finds that the fragmentation structure of the TL model under dephasing noise is determined by $\mathcal{C}_{\mathrm{PF}}$.

\subsection{Derivation of non-equilibrium stationary states}\label{subapp:stationary_states}
With the analysis of the commutant algebra of the Lindblad system, we obtained a full set of conserved projectors $\Pi_{\alpha}$, which decompose the operator space into minimal subspaces with independent dynamics. 

Now we prove the uniqueness of the eigenstates with zero eigenvalues (fixed points) of the Liouvillian within the minimal subspaces. This can be explained as follows \cite{Baumgartner_2008, Baumgartner_2008_2}: Density matrices form a convex set $\mathcal{S}$, where the boundary $\partial \mathcal{S}$ consists of all states with a lower rank~\cite{Bengtsson_2012_bloch_sphere}. Assume that both $\rho^1_\infty$ and $\rho_\infty^2$ are stationary states in one diagonal minimal block $\mathcal{B}_{\alpha}$. Due to the linearity of the Lindblad equation, $\rho_\infty^\lambda = \lambda \rho^1_\infty + (1-\lambda) \rho^2_\infty$ forms a line of stationary states. Assume that the line intersects with the boundary $\partial S$ at $\rho^3_\infty$, which has  $\mathrm{rank} (\rho_\infty^3)$ smaller than the dimension of the subspace. 
The range of $\rho^3_\infty$ is then a smaller subspace that contains a stationary state. This indicates that we can further decompose $\mathcal{B}_\alpha$, which is a contradiction with the fact that $\mathcal{B}_\alpha$ is a minimal subspace. Therefore, within each diagonal minimal block, there is at most one stationary state. In our case, we have proven that the unique stationary state within the subspace is the projected identity, $\Pi_{\alpha}\mathbb{1}/D_\alpha \in \mathcal{B}_{\alpha}$. 

In the off-diagonal subspaces, the existence of fixed points, i.e., the stationary coherences, is guaranteed by the conserved intertwined operators $\Pi_{\alpha \alpha^\prime}^{\lambda}$ between two degenerate diagonal subspaces for non-Abelian commutant algebras. This is given in Theorem 18 of Ref.~\cite{Baumgartner_2008_2}. There is also a unique fixed point in each off-diagonal subspace. Otherwise, the conserved intertwine operator gives an extra fixed point in the diagonal subspace, which is a contradiction.

To obtain the general expression of the stationary state in the full operator space, we perform a spectral decomposition of the Liouvillian superoperator. Due to the non-Hermiticity of $\mathcal{L}$, there is in general a different set of eigenstates for $\mathcal{L}$ and $\mathcal{L}^\dagger$ given by \cite{2020_intro_Lindblad, 2014_Albert}
\begin{equation}
    \mathcal{L} (\rho_n) = \lambda_n \rho_n,\,\,\,\,\, \mathcal{L}^\dagger (\sigma_n) = \lambda^*_n \sigma_n, 
\end{equation}
or equivalently $(\sigma^\dagger_n)\mathcal{L}=\lambda_n \sigma_n^\dagger$ for the latter, i.e.,  different left and right eigenstates of $\mathcal{L}$.
They satisfy the biorthonormal relation $\langle\langle \sigma_m |\rho_n \rangle\rangle \equiv \mathrm{Tr}(\sigma_m^\dagger \rho_n) = \delta_{mn}$. The left and right eigenmatrices span a full basis, such that we can expand the initial state as $\rho_0 = \sum_n c_n \rho_n$, with $c_n = \mathrm{Tr}(\sigma_n^\dagger \rho_0)$. The eigenspectrum of $\mathcal{L}$ consists of eigenvalues with $\mathrm{Re}(\lambda)\leq 0$. Consider systems without purely-imaginary eigenvalues: in the long time limit, the dynamics is governed by eigenmatrices with zero eigenvalues. Therefore, the full stationary state is then given by \cite{2014_Albert}
\begin{equation}\label{eq:general_stationary_state}
    \rho_{\text{ss}} = \lim_{t\rightarrow\infty} e^{t\mathcal{L}}(\rho_0) = \sum_n \mathrm{Tr}(\sigma_n^\dagger \rho_0) \rho_n,  
\end{equation}
with $\mathcal{L}^\dagger(\sigma_n) = 0$ and $\mathcal{L}(\rho_n) = 0 $.

With the analysis of the Hilbert space structure, we have identified the full set of eigenstates with zero eigenvalues of $\mathcal{L}$, which are the stationary states and stationary coherences $\{\Pi_{\alpha \alpha^\prime}\mathbb{1}\}$, as well as their corresponding conserved quantities $\{\Pi_{\alpha\alpha^\prime}\}$. With Eq.~\eqref{eq:general_stationary_state}, we obtain the general expression for the stationary state of fragmented systems, specified in the main text by Eq.~\eqref{eq:rho_ss_classical} and Eq.~\eqref{eq:rho_ss_quantum} in the case of classical and quantum fragmentation respectively. 

\subsection{Mazur bound in open system}\label{subapp:Mazurbound_open}
The Mazur bound in closed systems relates the infinite-time average value of autocorrelation functions (non-ergodicity) to the presence of conserved quantities~\cite{MAZUR1969, SUZUKI1971, Mazur_2021}. In the main text we derived the saturation value of autocorrelation function $\langle O(\infty) O(0)\rangle$ by the stationary value of the operator $O$, which coincides with the Mazur bound of the closed systems. The same conclusion can be achieved by generalizing the Mazur bound to open systems. 

We consider a diagonalizable Liouvillian $\mathcal{L}$~\cite{2014_Albert} with strong symmetries. The vectorized notation of the Liouvillian is~\cite{2020_intro_Lindblad}
\begin{equation}
        \mathcal{L}\rho \rightarrow \tilde{\mathcal{L}}|\rho\rangle\rangle, \ 
        \mathrm{Tr}(A^\dagger B) \rightarrow \langle\langle A|B\rangle\rangle,
\end{equation}
where the density operator is represented as a state and the Liouvillian becomes an operator in the Fock-Liouvillian space.

For a finite system and a diagonalizable $\mathcal{L}$, the left and right eigenmatrices $\{l_i^\dagger\}$ and $\{r_i\}$ form a complete set of basis, i.e. $\sum_i |r_i\rangle\rangle \langle \langle l_i| =0$ and satisfy the biorthonormal relation.
Thus we can expand arbitrary observable $Y$ as 
$|Y\rangle\rangle=\sum_{i}|r_{i}\rangle\rangle\langle\langle l_{i}|Y\rangle\rangle$~\cite{2020_intro_Lindblad}.
The autocorrelation function is
\begin{equation}
\begin{aligned}
\langle Y(t)Y(0)\rangle & =\frac{1}{D}\langle\langle Y|e^{t\tilde{\mathcal{L}}}Y\rangle\rangle,\\
 & =\frac{1}{D}\sum_{ij}\langle\langle Y|r_{i}\rangle\rangle\langle\langle l_{i}|e^{t\tilde{\mathcal{L}}}|r_{j}\rangle\rangle\langle\langle l_{j}|Y\rangle\rangle,\\
 & =\frac{1}{D}\sum_{ij}e^{\lambda_{j}t}\langle\langle Y|r_{i}\rangle\rangle\langle\langle l_{i}|r_{j}\rangle\rangle\langle\langle l_{j}|Y\rangle\rangle,\\
 & =\frac{1}{D}\sum_{j}e^{\lambda_{j}t}\langle\langle Y|r_{j}\rangle\rangle\langle\langle l_{j}|Y\rangle\rangle,
\end{aligned}
\end{equation}
with $D$ as the dimension of Hilbert space, and $Y=Y^\dagger$. 

With strong symmetries, the set of conserved quantities $\{J_\mu\}$ satisfy $[H, J_\mu] = [L_j, J_\mu] = 0$ for all $L_j$, indicating that $\tilde{\mathcal{L}}|J_\mu\rangle\rangle = 0$. Therefore, the corresponding right eigenmatrices are also given by $J_\mu$~\footnote{For general Liouvillian this is not necessarily the case~\cite{2014_Albert}.}. 

In the long-time average, all the oscillating ($\mathrm{Im} \lambda\neq 0$) and decaying terms ($\mathrm{Re}\lambda<0$) vanish. Therefore, the contributions will be given by the $\{J_{\mu}\}$ which are associated with zero eigenvalues,
\begin{equation}\label{eq:auto_non_neg}
\overline{\langle Y(t)Y(0)\rangle} = \frac{1}{D}\sum_{\mu}\frac{|\langle\langle Y|J_{\mu}\rangle\rangle|^2}{\langle\langle J_\mu | J_\mu \rangle\rangle},
\end{equation}
where we define $\overline{\langle Y(t) \rangle} \equiv \lim_{T\rightarrow\infty}\frac{1}{T}\int_{0}^{T} Y(t) dt$. This is the Mazur bound in the open system, with a set of orthogonal conserved quantities $\{J_\mu\}$. In general, for conserved quantities $\{Q_\mu\}$ not orthogonal, the Mazur bound can be written as
\begin{equation}
    M_A := \sum_{\mu\nu} \langle AQ_\mu\rangle \langle K^{-1}\rangle_{\mu\nu} \langle Q_\nu^\dagger A\rangle,
\end{equation}
where $(K)_{\mu\nu} \equiv \langle Q_\mu^\dagger Q_\nu\rangle$ is the correlation matrix. 

For open fragmented systems, with the choice of $L_j \in \mathcal{A}$, all elements in $\mathcal{C}$ commute with the Hamiltonian and jump operators. A full set of orthogonal conserved quantities in $\mathcal{C}$ are the projectors onto the Krylov subspaces $\{\Pi_{\alpha}\}$ and the intertwine operators $\{\Pi_{\alpha \alpha^\prime}\}$. With jump operators $S_j^z$ and $e_{j,j+1}$, the Mazur bound gives the same results derived from the stationary states in Eq.~\eqref{eq:Mazur_classical} and Eq.~\eqref{eq:Mazur_quantum}, respectively. 

\section{Effective Lindblad dynamics under dephasing noise}
\subsection{Perturbation theory of effective Lindblad dynamics}\label{subapp:SW_Lindblad}
We used the generalized Schrieffer-Wolff transformation to derive the effective Lindblad dynamics under large dephasing noise \cite{2012SW_for_dissipation}. The Liouvillian superoperator can be partitioned into $\mathcal{L} = \mathcal{L}_0 + \mathcal{L}_1$, with
\begin{equation}
\begin{aligned}
    \mathcal{L}_0 (\rho) &= \gamma \sum_j \left(S_j^z \rho S_j^z - \frac{1}{2}\{(S_j^z)^2, \rho\}\right),\\
    \mathcal{L}_1 (\rho) &= -i[\sum_j J_j e_{j,j+1}, \rho].
\end{aligned}
\end{equation}
In the $|J_j|/\gamma \rightarrow 0$ limit, the (right) eigenspectrum of the Liouvillian has a spectral gap between $\lambda_0 = 0$ and $\{\lambda_i, i>0, \lambda_i \neq 0\}$. Let the projector $\mathcal{P}$ onto the stationary state subspace spanned by eigenmatrices with $\lambda_0 = 0$. To second order in $|J_j|/\gamma$, perturbation theory gives $\mathcal{L}_{\text{eff}} = \mathcal{P}\mathcal{L}_1(\lambda_0 - \mathcal{L}_0)^{-1}\mathcal{L}_1\mathcal{P}$~\cite{2012SW_for_dissipation}.

For the initial state $|\psi_0\rangle = \otimes_{j} |+\rangle_j$, the stationary state subspace is spanned by $\rho_0^{\boldsymbol{\sigma}} = |\boldsymbol{\sigma}\rangle\langle\boldsymbol{\sigma}|$, with $|\boldsymbol{\sigma}\rangle$ all the product states in the fully-paired (zero dots) subspace of the PF model. The intermediate states are given by $|\boldsymbol{\sigma}\rangle \langle \boldsymbol{\sigma}'|$, with $|\boldsymbol{\sigma'}\rangle \equiv |\boldsymbol{\sigma}, j, \alpha,\beta\rangle = (|\beta\beta\rangle\langle\alpha\alpha|)_{j,j+1}|\boldsymbol{\sigma}\rangle$ and $\alpha\neq\beta$. The unperturbed Liouvillian $\mathcal{L}_0$ acting on the intermediate states gives eigenvalues $-4\gamma$ whenever $\alpha\beta\neq0$ or $-\gamma$ when $\alpha\beta=0$. Therefore, the effective Liouvillian acting on the stationary state subspace reduces to a classical stochastic evolution
\begin{equation}\label{eq:eff_Liouvillian_dephasing}
    \mathcal{L}_{\text{eff}} (\rho_0^{\boldsymbol{\sigma}}) = -\sum_{\boldsymbol{\sigma}'}\langle \boldsymbol{\sigma}'|\mathbb{W}_{\text{eff}}|\boldsymbol{\sigma}\rangle\rho_0^{\boldsymbol{\sigma}'}.
\end{equation}
The effective Markov generator reads $\mathbb{W}_{\text{eff}}=\sum_{i}\frac{J_{i}^{2}}{\gamma} M_{i,i+1}$, with matrix representation 
\begin{equation}
M_{i,i+1}=\text{\ensuremath{\frac{1}{2}}}\begin{pmatrix}5 &  &  &  & -4 &  &  &  & -1\\
 & 0\\
 &  & 0\\
 &  &  & 0\\
-4 &  &  &  & 8 &  &  &  & -4\\
 &  &  &  &  & 0\\
 &  &  &  &  &  & 0\\
 &  &  &  &  &  &  & 0\\
-1 &  &  &  & -4 &  &  &  & 5
\end{pmatrix},
\end{equation}
in the local $z$ basis for two consecutive sites.
The dynamics given by the matrix $\mathbb{W}_{\text{eff}}$ has the same block-diagonal structure than the PF model in the local $z$ basis, $H_{\text{PF}} = \sum_i g_i^{\alpha\beta} (|\alpha\alpha\rangle\langle\beta\beta|)_{i,i+1}$,
with $g_i^{\alpha\beta} = \frac{J_i^2}{\gamma} \langle \alpha \alpha |M_{i,i+1}|\beta\beta\rangle$. This indicates that effective dynamics of the TL model under dephasing noise is given by the PF model. 

\subsection{Simulation of the stochastic dynamics}\label{subapp:stochastic_dynamics}
The effective dynamics of the TL model under dephasing noise can be mapped to a classical Markov process, with the stochastic matrix $\mathbb{W}_{\text{eff}}$ in Eq.~\eqref{eq:eff_Liouvillian_dephasing}. Here we provide more details of our numerical simulation. In correspondence to the random circuit setting for the quantum Lindblad dynamics, we implement the dynamics by classical circuits, with two-site gates $U_{t,j}$ permuting among the classical configurations $\boldsymbol{\sigma} = \{s_1, ..., s_N\}$, where $s_{i} \in \{+, 0, -\}$. These classical configurations $\boldsymbol{\sigma}$ correspond to $\rho_0^{\boldsymbol{\sigma}}$ with $|\boldsymbol{\sigma}\rangle = |\sigma_1, ..., \sigma_N\rangle$. A two-site gate $U_{t,j}$ acting on the configuration $\boldsymbol{\sigma} = \{..., s_j, s_{j+1}, ...\}$ at time $t$ gives a new configuration $\boldsymbol{\sigma}' = \{..., s'_j, s_{j+1}', ...\}$ with a transition probability. A configuration with $s_j = s_{j+1}$ can transform to a new configuration with $s_j' = s_{j+1'}'$, with the transition probability given by the probability matrix $\langle s_j' s_{j+1}'|P_{j,j+1}|s_j s_{j+1}\rangle$ with $P_{j,j+1} = \exp(-\frac{J_j^2}{\gamma}M_{i,i+1})$~\cite{2016_Dephasing_MBL}. This is a pair-flip action. For $s_j\neq s_{j+1}$, the configuration is unchanged. To compare with the Lindblad dynamics, we start from the initial configuration with all $s_i = +$. Averaging over random circuit realizations, we obtain the time evolution of the probability of dot patterns $p_{A_k} (t)$, and thus the number entanglement $S_{\text{num}}(t)$. The mapping of the effective dynamics to the stochastic dynamics allows simulation for much larger system sizes and longer times. 

\subsection{Saturation of OSE}\label{subapp:Sat_OSE}
For the initial state $|\psi_0\rangle = \otimes_j |+\rangle$ which lies within in the fully-paired subspace, the stationary state is the corresponding projected identity $\rho_{\text{ss}}\propto \Pi_{\mathcal{K}_{\text{paired}}} \mathbb{1} = \sum_{|\boldsymbol{\sigma}\rangle \in \mathcal{K}_{\text{paired}}}|\boldsymbol{\sigma}\rangle\langle\boldsymbol{\sigma}|$, where $|\boldsymbol{\sigma}\rangle$ are the product basis states of the fully-paired subspace, $|\boldsymbol{\sigma}\rangle = |\sigma_1, \sigma_2, ...\rangle$. By vectorizing $|\sigma_1, \sigma_2, ...\rangle\langle \sigma_1, \sigma_2, ...| \rightarrow |\sigma_1\sigma_1, \sigma_2\sigma_2, ...\rangle\rangle$, we can map the stationary mixed state to a pure state $|\psi_{\text{ss}}\rangle \propto \sum_{|\boldsymbol{\sigma}\rangle\in\mathcal{K}_{\text{paired}}} |\boldsymbol{\sigma}\rangle$, which is an equal superposition of the fully dimerized states. The bipartite OSE of the stationary state equals to the bipartite entanglement entropy of the vectorized state. 

The half-chain entanglement of $|\psi_{\text{ss}}\rangle$ was studied in Ref.~\cite{2018_PFmodel}. It was shown that the symmetry-resolved entanglement $S_{\text{res}} = \sum_{A_k} p_{A_k} S_{\text{res}}(A_k)= 0$, as after resolving the left dot pattern $A_{k}$, all the configurations contribute equally (thus the state can be written as a product state with $S_{\text{res}}(A_k) = 0$ for all $A_k$. Hence, only the number entropy given by $S = -\sum_{A_k} p_{A_k} \log p_{A_k}$, with $A_k$ as the dot pattern of the left part of the chain remains.  For large system sizes $N$, the entanglement scales as $S \sim O(\sqrt{N})$. 

\begin{figure}[tb]
	\centering
	\includegraphics[width=6cm]{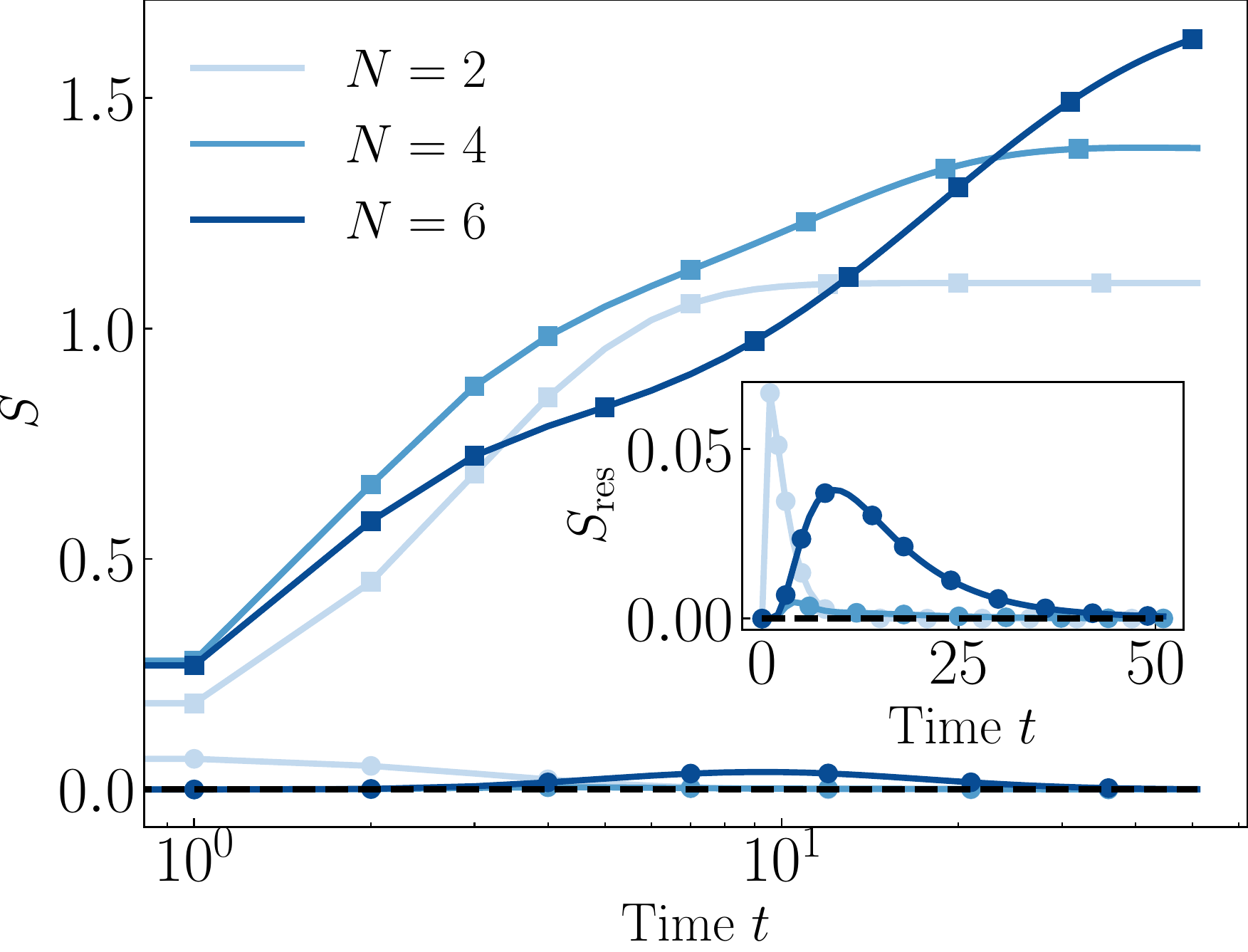}
	\caption{\textbf{Number entanglement and symmetry-resolved entanglement.} Time evolution of the number entanglement $S_{\text{num}}$ (square) and the symmetry-resolved entanglement $S_{\text{res}}$ (circle) for different system sizes $N$ by ED. The $S_{\text{res}}$ is small compared to $S_{\text{num}}$, and saturates to zero.}
	\label{fig:TL3_dephasing_S_num_S_res}
\end{figure}

In Fig.~\ref{fig:TL3_dephasing_S_num_S_res}, we show the Lindblad evolution of $S_{\text{num}}$ and $S_{\text{res}}$ under dephasing noise for $\gamma = 10$. We show that the symmetry-resolved entanglement is small compared to the number entanglement during time evolution, and saturates to zero for the stationary states. 

\providecommand{\noopsort}[1]{}\providecommand{\singleletter}[1]{#1}%

\end{document}